\documentclass[prd,aps,twocolumn,tightenlines,superscriptaddress,floatfix]{revtex4-2}

\usepackage{newtxtext,newtxmath}
\usepackage[T1]{fontenc}
\usepackage[english]{babel}
\usepackage{booktabs}
\usepackage{tabularx}
\usepackage{graphicx}
\usepackage{amsmath}
\usepackage{bm}
\usepackage{xcolor}
\usepackage{hyperref}

\hypersetup{colorlinks=true,linkcolor=blue!55!black,urlcolor=blue!55!black,citecolor=blue!55!black}

\begin{document}

\title{Milky Way Structure from Double White Dwarf Gravitational-Wave Sources}

\author{Shao-Dong Zhao}
\affiliation{Institute of Theoretical Physics \& Research Center of Gravitation, Lanzhou University, Lanzhou 730000, China}
\affiliation{Lanzhou Center for Theoretical Physics, Key Laboratory of Theoretical Physics of Gansu Province, Key Laboratory of Quantum Theory and Applications of MoE, Gansu Provincial Research Center for Basic Disciplines of Quantum Physics, Lanzhou University, Lanzhou 730000, China}

\author{Xue-Hao Zhang}
\affiliation{Institute of Theoretical Physics \& Research Center of Gravitation, Lanzhou University, Lanzhou 730000, China}
\affiliation{Lanzhou Center for Theoretical Physics, Key Laboratory of Theoretical Physics of Gansu Province, Key Laboratory of Quantum Theory and Applications of MoE, Gansu Provincial Research Center for Basic Disciplines of Quantum Physics, Lanzhou University, Lanzhou 730000, China}
\affiliation{School of Fundamental Physics and Mathematical Sciences, Hangzhou Institute for Advanced Study, UCAS, Hangzhou 310024, China}

\author{Qun-Ying Xie}
\affiliation{School of Information Science \& Engineering, Lanzhou University, Lanzhou 730000, China}

\author{Yu-Xiao Liu}
\affiliation{Institute of Theoretical Physics \& Research Center of Gravitation, Lanzhou University, Lanzhou 730000, China}
\affiliation{Lanzhou Center for Theoretical Physics, Key Laboratory of Theoretical Physics of Gansu Province, Key Laboratory of Quantum Theory and Applications of MoE, Gansu Provincial Research Center for Basic Disciplines of Quantum Physics, Lanzhou University, Lanzhou 730000, China}

\author{Soumya D. Mohanty}
\email{Soumya.mohanty@utrgv.edu}
\affiliation{Department of Physics and Astronomy, University of Texas Rio Grande Valley, One West University Blvd., Brownsville, Texas 78520, USA}

\date{\today}

\begin{abstract}
The millihertz gravitational-wave sky will be dominated by $\sim10^{8}$ double white dwarfs while $\sim10^{4}$ be resolvable in the Milky Way, whose three-dimensional spatial distribution traces the Galaxy's structural parameters.  We present \textsc{Galena}, a hierarchical Bayesian pipeline that recovers these parameters from a double white dwarf catalogue through an inhomogeneous Poisson-process likelihood.  Each source's $3\times3$ Galactic position covariance is computed from the waveform Fisher information matrix and propagated analytically through a $3\times5$ Jacobian; the resulting measurement-error convolution is factored into a pre-computed sparse weight matrix, rendering the nested-sampling inference tractable.  Applied to GBSIEVER-reported LISA Data Challenge sources ($N=2151$ after quality cuts), \textsc{Galena} recovers the disk scale length $R_d=2175^{+51}_{-52}$~pc and scale height $z_d=282^{+7}_{-7}$~pc, consistent with a canonical thin disk, together with the bulge fraction $A=0.187^{+0.011}_{-0.012}$ and bulge scale radius $R_b=773^{+26}_{-25}$~pc; the bulge fraction, now recovered much closer to the literature value of $0.25$ than in our earlier exact-position fit, is measured at $\sim6\%$ statistical precision.  An independent particle-swarm optimisation of the same likelihood reproduces these values to within a fraction of a percent, supporting the attribution of the improvement to the Fisher-matrix error propagation rather than to the sampler.  The search-stage astrophysical prior improves per-source distance estimates but leaves the hierarchical inference essentially unchanged.
\end{abstract}


\maketitle

\section{Introduction}

The 2030s will witness the opening of the millihertz gravitational-wave (GW) window.  Three space-based detectors, LISA \cite{AmaroSeoane2017}, Taiji \cite{Hu2017,ruan2020taiji}, and TianQin \cite{Luo2016}, will survey the $10^{-4}$--$10^{-1}$~Hz frequency band with sensitivity sufficient to resolve tens of thousands of individual Galactic sources.  Among these, double white dwarf (DWD) binaries are by far the most numerous: population-synthesis models predict $\sim10^{4}$ resolvable DWDs in the Milky Way, forming a confusion-limited foreground that encodes a detailed map of the Galaxy's stellar mass distribution \cite{Nelemans2001,Ruiter2010,Korol2022,karnesis2021stochastic,korol2020populations}.

Each resolved DWD is a tracer of Galactic structure.  Because the spatial distribution of DWDs follows the underlying stellar populations (thin and thick disks, bulge, and halo), the observed catalogue of positions and distances carries information about the structural parameters of the components modelled here: thin disk scale length $R_{d}$ and scale height $z_{d}$, bulge fraction $A$ and scale radius $R_{b}$ of bulge.  Extracting these parameters from a noisy, incomplete catalogue is a hierarchical Bayesian inference problem that has attracted growing attention over the past decade \cite{breivik2020constraining,korol2019multimessenger,Wilhelm:2020qjc,georgousi2023gravitational,zhang2024constraining}.

The idea of using resolved DWDs to map the Milky Way goes back to \cite{Littenberg2011}.  \cite{Wilhelm:2020qjc} extended it to the Galactic bar and spiral arms, reconstructing a simulated Milky Way from mock LISA observations of a realistic DWD population embedded in a high-resolution $N$-body simulation: the bar appeared clearly in the GW map of the bulge, with its viewing angle recovered to within $\sim1^\circ$ and its axis ratio to within $1\sigma$, while the spiral arms could not be characterised.  Our own earlier work began turning this idea into practice.  In \cite{zhang2021pso} we developed the GBSIEVER search pipeline, a particle-swarm-optimisation (PSO) \cite{eberhart1995particle} search with cross-validation that resolves individual Galactic binaries from the LISA data stream, and in \cite{universe11080248} we introduced a two-component (bulge $+$ thin disk) density model and estimated the Galactic structural parameters with a maximum-likelihood fit, showing that $\sim10^{3}$ resolved DWDs could yield disk parameters consistent with literature, unlike the bulge parameters, which show clear deviations. Although $R_{d}$ and $z_{d}$ were recovered at roughly the expected precision, that fit suffered from two recognised limitations, the first of which this paper resolves, while the second is only formalised within the IPP framework, leaving a realistic treatment to future work.

The fit neglected measurement uncertainty: DWD distances are not observed directly but are inferred from the gravitational-wave frequency, its time derivative, and the strain amplitude through the quadrupole formula, and the resulting errors \cite{balasubramanian1996gravitational}---chiefly in $\dot{f}$ and $\mathcal{A}$---translate into fractional distance uncertainties of $\sigma_{R}/R\sim15$--$50\%$ at the signal-to-noise ratios (SNR) typical of LISA sources (SNR$\sim10$--$100$), i.e. absolute position errors of $\sim0.5$--$3$~kpc \cite{Korol2022}.  Because this uncertainty is much larger than the disk scale height ($z_{d}\sim300$~pc), the vertical structure of the thin disk is effectively smeared out, leaving the bulge fraction $A$ nearly degenerate with the disk parameters---an effect clearly visible in our earlier maximum-likelihood fit, which failed to constrain $A$ at all.

In gravitational-wave astronomy the Fisher-matrix formalism is the standard tool for characterising measurement errors: the Fisher information matrix---the noise-weighted inner products of the signal derivatives with respect to its parameters---approximates the inverse parameter covariance at high signal-to-noise ratio.  In the present context the crucial step is to propagate this per-source parameter covariance into the three-dimensional Galactic position, which is a nonlinear function of the waveform parameters through the distance relation (Eq.~\ref{eq:distance}); this propagation is what couples the per-source measurement model to the Galactic density model.  We compute the Fisher matrix directly from the time-delay-interferometry response of the detector, and marginalise over the inclination (which is degenerate with the amplitude); the resulting position covariances are then convolved into the likelihood through the pre-computed weight matrix of Sec.~\ref{sec:sparse_W}.

The selection function was handled only approximately.  A signal-to-noise-limited survey preferentially detects louder sources, which tend to be closer or at higher frequency, so a density model fitted without correcting for this detection bias develops Malmquist-like distortions: distant sources are under-represented in the catalogue, and the fitted scale length $R_{d}$ is consequently biased downward, as the model steepens the density profile to match the apparent deficit of distant sources.  That earlier fit instead relied on a hard distance cut ($R_{\rm max}=15$~kpc), a crude but effective first step; a realistic selection function $\bar{\eta}(\mathcal{M}_{c},f,R)$ built from end-to-end injection-recovery simulations remains an important goal for future work.

The key difference from \cite{universe11080248} is therefore the likelihood: we replace its exact-position maximum-likelihood objective with an IPP likelihood that convolves per-source position uncertainties built from the waveform Fisher matrix.  Sampling this posterior with nested sampling (\texttt{dynesty}) rather than maximising it with PSO is a convenience, not the source of the improvement; indeed, the two methods' best-fit parameters agree to within a fraction of a percent (Sec.~\ref{sec:results}), which is consistent with the gain tracing to the likelihood itself rather than to the sampler.

In this paper we present \textsc{Galena} and apply it to GBSIEVER search outputs for the LISA data challenge (LDC) and Taiji data challenge (TDC I) data streams.  TianQin is not included in the present study, because no public TianQin mock-data catalogue or data challenge comparable to LDC and TDC has yet been released; since the pipeline is detector-agnostic, it can be applied to TianQin data as soon as such a catalogue becomes available.  Our contributions are threefold.  (i) \emph{Methodology:} we propagate per-source GW parameter uncertainties analytically into full $3\times3$ Galactic position covariances via a $3\times5$ Jacobian (Sec.~\ref{sec:jacobian}), and factor the resulting measurement-error convolution out of the density-model integration into a pre-computed sparse weight matrix (Sec.~\ref{sec:sparse_W}), rendering full Bayesian inference with an inhomogeneous Poisson process (IPP) likelihood tractable for catalogues of thousands of sources.  (ii) \emph{First search-pipeline results:} applying the pipeline to the GBSIEVER-reported LDC catalogue, we recover the disk scale parameters to a few per cent and obtain a few-percent constraint on the bulge fraction $A$, which our earlier maximum-likelihood fit could not constrain at all.  (iii) \emph{Quantitative diagnostics:} we characterise the effect of the astrophysical prior used in the GBSIEVER search stage on the source frequency derivative, and establish the grid resolution required for unbiased recovery of sub-kiloparsec structure.

The remainder of this paper is organised as follows.  Sec.~\ref{sec:method} develops the methodological framework: the Galactic density model (Sec.~\ref{sec:density}), the analytical mapping from GW parameters to 3D Galactic positions with full Jacobian-based error propagation (Sec.~\ref{sec:jacobian}), the IPP likelihood with a selection-function term (Sec.~\ref{sec:ipp}), and the sparse pre-computation strategy that makes the inference computationally tractable (Sec.~\ref{sec:sparse_W}).  Sec.~\ref{sec:data} describes the catalogues used in this analysis.  Sec.~\ref{sec:results} validates the pipeline on synthetic data and then presents the main results: posterior constraints from GBSIEVER-reported LDC and TDC sources, a comparison with the LDC truth catalogue, the impact of the astrophysical prior used in the search stage, and grid-resolution convergence tests.  Sec.~\ref{sec:conclusions} summarises the main results, discusses the limitations of the current method, and outlines directions for future work.

\section{Method}
\label{sec:method}

\begin{figure}[t]
\centering
\includegraphics[width=\columnwidth]{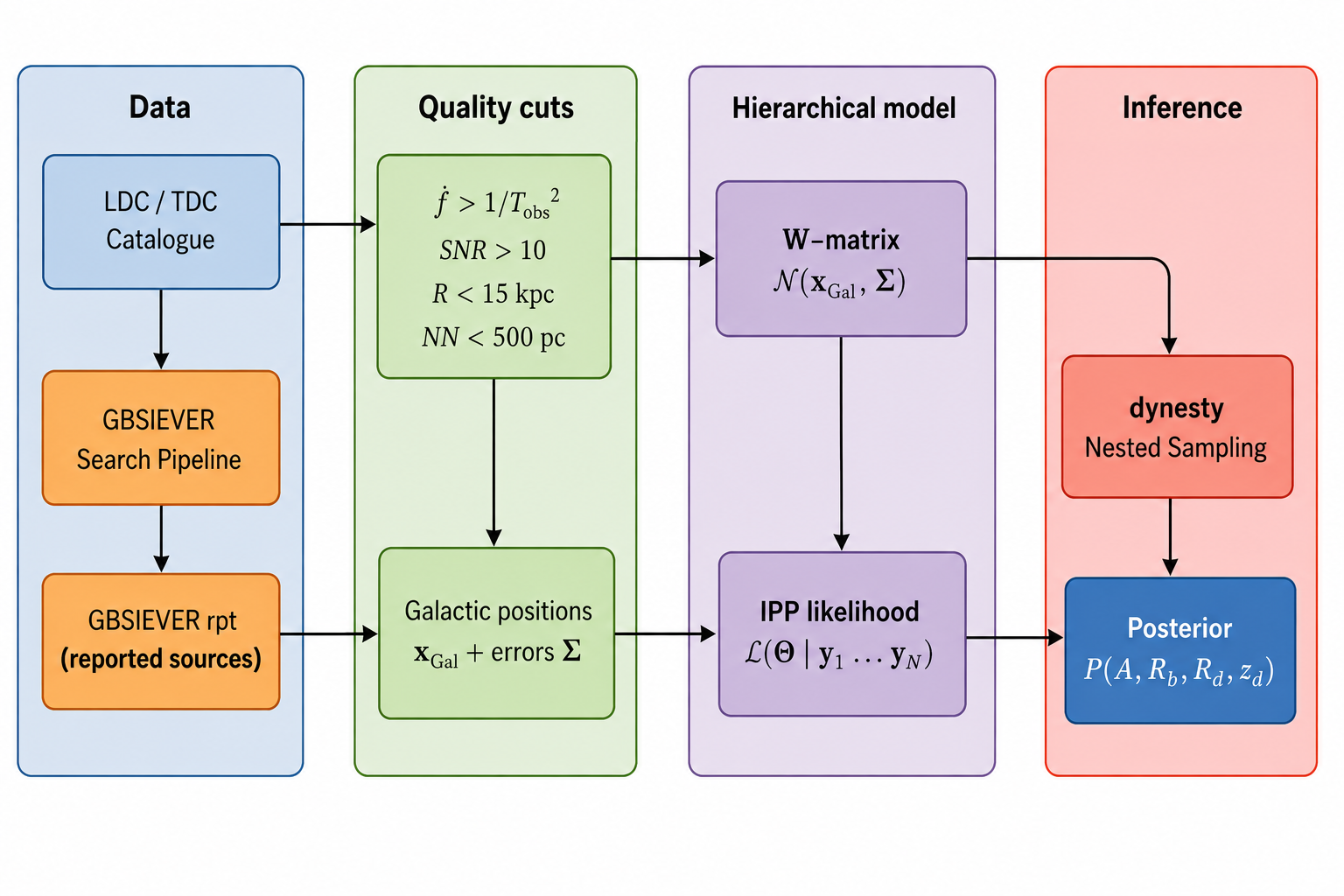}
\caption{Schematic of the \textsc{Galena} pipeline: search-catalogue parameters and SNRs are propagated to 3D Galactic positions with Jacobian-based error propagation (Sec.~\ref{sec:jacobian}); the IPP likelihood with a pre-computed sparse weight matrix (Secs.~\ref{sec:ipp}--\ref{sec:sparse_W}) links the observed catalogue to the structural parameters $\Theta$, which are sampled with \texttt{dynesty}~\protect\cite{Speagle2020}.}
\label{fig:method}
\end{figure}

This section develops the method in five parts: the DWD signal model, which defines the eight waveform parameters and the $\mathcal{F}$-statistic used to extract them (Sec.~\ref{sec:signal}); the two-component Galactic density model, whose four parameters $\Theta=(A,R_b,R_d,z_d)$ are to be inferred (Sec.~\ref{sec:density}); the mapping from GW parameters to three-dimensional Galactic positions with full Jacobian-based error propagation (Sec.~\ref{sec:jacobian}); the inhomogeneous Poisson-process likelihood that folds the per-source errors into the density model (Sec.~\ref{sec:ipp}); and the sparse weight-matrix pre-computation that renders the inference tractable (Sec.~\ref{sec:sparse_W}).  

Fig.~\ref{fig:method} summarises \textsc{Galena} (\textbf{GAL}actic \textbf{E}stimation with \textbf{N}ested s\textbf{A}mpling), our hierarchical Bayesian inference pipeline.  Per-source GW parameters and their uncertainties from the search catalogues (Sec.~\ref{sec:data}) are mapped to 3D Galactic positions with full analytical error propagation (Sec.~\ref{sec:jacobian}); an IPP likelihood with a pre-computed sparse weight matrix (Sec.~\ref{sec:ipp}, Sec.~\ref{sec:sparse_W}) then connects the observed catalogue to the Galactic structure parameters, which are explored with nested sampling.

\subsection{DWD signal model}
\label{sec:signal}

A DWD in a circular orbit emits an almost monochromatic gravitational-wave signal whose frequency slowly increases as the orbit decays under gravitational radiation.  In the quadrupole approximation the signal observed by a space-based detector is described by eight parameters, split into four intrinsic parameters $\kappa=(f,\dot{f},\beta,\lambda)$---the gravitational-wave frequency $f$ and its time derivative $\dot{f}$, together with the ecliptic latitude $\beta$ and longitude $\lambda$---and four extrinsic parameters---the dimensionless strain amplitude $\mathcal{A}$, the polarisation $\psi$, the inclination $\iota$, and the initial phase $\varphi_{0}$.  Because the phase evolution and the sky-response modulation depend only on the intrinsic parameters, the $\mathcal{F}$-statistic \cite{Jaranowski1998} reduces the data to a function of $\kappa$ by maximising analytically over the extrinsic parameters,
\begin{equation}
\mathcal{F}(\kappa)=\mathbf{U}(\kappa)^{\mathsf{T}}\mathbf{W}^{-1}\mathbf{U}(\kappa),
\label{eq:fstat}
\end{equation}
where $\mathbf{U}(\kappa)$ collects the noise-weighted correlations between the data and the four template waveforms and $\mathbf{W}$ is the template--template correlation matrix \cite{zhang2021pso}.  The intrinsic parameters are estimated by maximising $\mathcal{F}(\kappa)$, after which the extrinsic parameters, including the amplitude $\mathcal{A}$, are recovered analytically from the maximum-likelihood solution.  The search pipeline used here (GBSIEVER, Sec.~\ref{sec:data}) performs this maximisation over $\kappa$ iteratively with particle-swarm optimisation and returns, per resolved source, the maximum-likelihood estimates of the eight parameters $\bm{\theta}=(f,\dot{f},\beta,\lambda,\mathcal{A},\psi,\iota,\varphi_0)$ together with a signal-to-noise ratio computed from the likelihood.

\subsection{Galactic density model}
\label{sec:density}

The spatial distribution of resolved DWDs traces the underlying stellar mass of the Milky Way.  To extract structural parameters from this distribution, we require a parametric model $\rho(\mathbf{x}\mid\Theta)$ that captures the dominant Galactic components while remaining computationally tractable for inference.  Following \cite{adams2012astrophysical}, we adopt a two-component model consisting of a spherical bulge and an axisymmetric thin disk.

The bulge is modelled as a spherical Gaussian, motivated by the approximately spheroidal morphology of the inner Galaxy.  The thin disk, where the majority of DWDs reside, follows an exponential radial profile with a $\operatorname{sech}^{2}$ vertical fall-off, a standard parameterisation for stellar disks that naturally reproduces the isothermal-sheet solution at large $z$:
\begin{align}
\rho_{\rm bulge}(\mathbf{x}\mid\Theta)
   &= \frac{1}{(\sqrt{\pi}\,R_{b})^{3}}\,
      \exp\!\left(-\frac{R^{2}+z^{2}}{R_{b}^{2}}\right),
      \label{eq:rho_bulge}\\[4pt]
\rho_{\rm disk}(R,z\mid\Theta)
   &= \frac{1}{4\pi R_{d}^{2}\,z_{d}}\,
      \exp\!\left(-\frac{R}{R_{d}}\right)
      \nonumber\\
   &\qquad\times \operatorname{sech}^{2}\!\left(\frac{z}{z_{d}}\right),
      \label{eq:rho_disk}
\end{align}
where $(R,z)$ are the usual Galactocentric cylindrical coordinates with $R\equiv\sqrt{x^{2}+y^{2}}$.  Both components are individually normalised to unit integral over $\mathbb{R}^{3}$, so that $\rho(\mathbf{x}\mid\Theta)\,d^{3}x$ is a proper probability density.  The full model is a convex combination
\begin{equation}
\rho(\mathbf{x}\mid\Theta)
   = A\,\rho_{\rm bulge}(\mathbf{x}\mid\Theta)
     + (1-A)\,\rho_{\rm disk}(\mathbf{x}\mid\Theta),
\label{eq:rho_total}
\end{equation}
controlled by the four-component parameter vector
\begin{equation}
\Theta = \bigl(A,\;R_{b},\;R_{d},\;z_{d}\bigr),
\end{equation}
where $A\in[0,1]$ is the dimensionless bulge fraction, $R_{b}$ the bulge scale radius, $R_{d}$ the disk scale radius, and $z_{d}$ the disk scale height (all in pc).  This model deliberately omits a thick disk and a stellar halo: the measurement uncertainties of current DWD catalogues (Sec.~\ref{sec:results}) are too large to resolve these sub-dominant components, and including them would weaken the already marginal constraints on the thin-disk parameters.

The choice of a two-component model is a compromise between astrophysical fidelity and statistical identifiability.  With four free parameters and $\sim10^{3}$--$10^{4}$ sources, the data volume is sufficient to constrain $R_{d}$ and $z_{d}$ but only marginally informative on $A$ and $R_{b}$, as we discuss in Sec.~\ref{sec:results}. A possible future refinement is to first separate the bulge and thin-disk components and then fit the parameters of each component independently.

\subsection{From GW parameters to 3D positions}
\label{sec:jacobian}

Of the eight parameters that describe a DWD signal (Sec.~\ref{sec:signal}), only five affect its three-dimensional position; the remaining extrinsic parameters ($\psi$, $\iota$, $\varphi_{0}$) do not.  These five are the frequency $f$, its time derivative $\dot{f}$, the ecliptic latitude $\beta$ and longitude $\lambda$, and the dimensionless strain amplitude $\mathcal{A}$, which we collect into the vector
\begin{equation}
\bm{\theta} = \bigl(f,\;\dot{f},\;\beta,\;\lambda,\;\mathcal{A}\bigr).
\end{equation}

\subsubsection{Distance and coordinates}

For a monochromatic binary, the GW frequency derivative is driven entirely by the orbital decay due to GW emission.  The luminosity distance is therefore expressible in terms of $f$, $\dot{f}$, and $\mathcal{A}$ alone \cite{Littenberg2011}:
\begin{equation}
R\bigl(\bm{\theta}\bigr)
   = \frac{5c}{48\pi^2}
     \frac{\dot{f}}{f^{3}\,\mathcal{A}},
\label{eq:distance}
\end{equation}
where the numerical prefactor follows from the quadrupole formula for a circular binary and $\mathcal{A}$ is the dimensionless strain amplitude in the convention $\mathcal{A}=h_0/2$ for the peak intrinsic amplitude $h_0$.

The solar-system-barycentre (SSB) ecliptic Cartesian position is obtained from the spherical coordinates $(R,\beta,\lambda)$:
\begin{align}
x_{\rm SSB} &= R \cos\beta \cos\lambda, \nonumber\\
y_{\rm SSB} &= R \cos\beta \sin\lambda, \nonumber\\
z_{\rm SSB} &= R \sin\beta.
\label{eq:ssb_xyz}
\end{align}
The transformation from the SSB ecliptic frame to the Galactic frame consists of a translation by the Sun--Galactic-Centre vector $\mathbf{r}_{\odot}$ followed by a rotation $\mathbf{R}_{\rm ecl\to gal}$:
\begin{equation}
\mathbf{x}_{\rm Gal}
   = \mathbf{R}_{\rm ecl\to gal}\,
     \bigl(\mathbf{x}_{\rm SSB} - \mathbf{r}_{\odot}\bigr).
\label{eq:ssb2gal}
\end{equation}
We adopt $\mathbf{r}_{\odot}=(-0.45,\,-8.45,\,-0.79)\;\mathrm{kpc}$ from the LDC catalogue, consistent with direct astrometric measurements of the Galactic centre \cite{reid2004proper,abuter2019geometric}, and extract $\mathbf{R}_{\rm ecl\to gal}$ numerically by applying the SSB-to-Galactic transformation to the standard basis vectors, ensuring consistency with the reference implementation.

\subsubsection{Position Jacobian}

Because $\mathbf{x}_{\rm Gal}$ is a deterministic, differentiable function of $\bm{\theta}$, we can compute its $3\times5$ Jacobian matrix analytically:
\begin{equation}
\mathbf{J}(\bm{\theta})
   \equiv \frac{\partial\mathbf{x}_{\rm Gal}}{\partial\bm{\theta}}
   = \mathbf{R}_{\rm ecl\to gal}\,
     \frac{\partial\mathbf{x}_{\rm SSB}}{\partial\bm{\theta}},
\label{eq:jacobian}
\end{equation}
where the rotation factorises out because it is parameter-independent. The SSB-frame Jacobian $\partial\mathbf{x}_{\rm SSB}/\partial\bm{\theta}$ follows from Eqs.~\eqref{eq:distance}--\eqref{eq:ssb_xyz} and evaluates to
\begin{align}
\frac{\partial\mathbf{x}_{\rm SSB}}{\partial f}
   &= -\frac{3}{f}\,\mathbf{x}_{\rm SSB}, \nonumber\\[2pt]
\frac{\partial\mathbf{x}_{\rm SSB}}{\partial\dot{f}}
   &=  \frac{1}{\dot{f}}\,\mathbf{x}_{\rm SSB}, \nonumber\\[2pt]
\frac{\partial\mathbf{x}_{\rm SSB}}{\partial\beta}
   &= \begin{pmatrix}
      -z_{\rm SSB}\,x_{\rm SSB}/\rho \\
      -z_{\rm SSB}\,y_{\rm SSB}/\rho \\
      \rho
      \end{pmatrix}, \nonumber\\[2pt]
\frac{\partial\mathbf{x}_{\rm SSB}}{\partial\lambda}
   &= \begin{pmatrix}
      -y_{\rm SSB} \\  x_{\rm SSB} \\ 0
      \end{pmatrix}, \nonumber\\[2pt]
\frac{\partial\mathbf{x}_{\rm SSB}}{\partial\mathcal{A}}
   &= -\frac{1}{\mathcal{A}}\,\mathbf{x}_{\rm SSB},
\label{eq:jac_ssb}
\end{align}
with $\rho\equiv\sqrt{x_{\rm SSB}^{2}+y_{\rm SSB}^{2}}$.  Near the ecliptic poles ($|\beta|\simeq\pi/2$), $\rho\to0$ produces a removable singularity in the $\beta$ column; we floor $\rho$ at $10^{-3}\,\mathrm{pc}$ to guarantee numerical stability without affecting the astrophysical region of interest.

\subsubsection{Waveform Fisher information matrix}
\label{sec:fim}

For the search catalogues we compute each source's Fisher matrix directly from the waveform model.  The DWD response is assembled in the TDI A/E/T channels, obtained from the three Michelson-like channels $(X,Y,Z)$ through the standard AET transformation; the per-channel response and its analytic derivatives with respect to the intrinsic parameters $(f,\dot{f},\beta,\lambda,\mathcal{A},\iota)$ are propagated through this transform, and the $6\times6$ Fisher matrix is formed as the noise-weighted time-domain inner product
\begin{equation}
F^{(i)}_{pq}
   = \sum_{\rm ch}
     \Bigl\langle \frac{\partial h_{\rm ch}}{\partial\theta_p},\,
                    \frac{\partial h_{\rm ch}}{\partial\theta_q}\Bigr\rangle,
\label{eq:fim}
\end{equation}
evaluated at the $i$th source's measured parameters with the per-channel one-sided noise power spectral density and the two-year observation time $T_{\rm obs}=62{,}914{,}560$~s of the LDC convention.  The inclination is retained in the Fisher matrix because it is degenerate with the amplitude: the two polarisation amplitudes scale as $\mathcal{A}(1+\cos^{2}\iota)/2$ and $\mathcal{A}\cos\iota$, so a source can trade amplitude against inclination without changing the signal.  Treating $\iota$ as a known parameter---equivalently, discarding its row and column of the $6\times6$ matrix---would therefore overestimate the precision on $\mathcal{A}$ and, through the distance scaling $R\propto\dot{f}/(f^{3}\mathcal{A})$ of Eq.~\eqref{eq:distance}, understate the per-source position error.  We instead marginalise $\iota$ out by the Schur complement~\cite{HAYNSWORTH196873}, inverting the $6\times6$ matrix and retaining the leading $5\times5$ block of its inverse, which is the $\mathbf{F}_{\bm{\theta}}^{(i)}$ over $\bm{\theta}=(f,\dot{f},\beta,\lambda,\mathcal{A})$ entering Eq.~\eqref{eq:fisher_prop}.

\subsubsection{Fisher propagation to position covariance}

Given per-source parameter uncertainties, encoded in full Fisher matrices $\mathbf{F}_{\bm{\theta}}^{(i)}\in\mathbb{R}^{5\times5}$ computed from the waveform model (Sec.~\ref{sec:fim}), the $3\times3$ position covariance in the Galactic frame follows from the standard error-propagation formula \cite{kay1993fundamentals}
\begin{equation}
\mathbf{C}_{\mathbf{x}}^{(i)}
   = \mathbf{J}\bigl(\bm{\theta}_i\bigr)\,
     \mathbf{F}_{\bm{\theta}}^{(i)\,{-1}}\,
     \mathbf{J}\bigl(\bm{\theta}_i\bigr)^{\!\mathsf{T}},
\label{eq:fisher_prop}
\end{equation}
where $\mathbf{J}(\bm{\theta}_i)$ is evaluated at the measured parameters of source $i$.  This $3\times3$ covariance fully captures the anisotropic and correlated position uncertainty produced by the non-linear mapping from GW parameter space to Galactic Cartesian space.

These covariances enter the IPP likelihood (Sec.~\ref{sec:ipp}) through the Gaussian measurement kernel that convolves the density model with each source's positional uncertainty.

\subsection{Inhomogeneous Poisson process likelihood}
\label{sec:ipp}

Given the density model $\rho(\mathbf{x}\mid\Theta)$ and the per-source position covariances $\mathbf{C}_{\mathbf{x}}^{(i)}$, we now construct a likelihood that connects the observed catalogue to the Galactic structure parameters.  The appropriate statistical framework is the inhomogeneous Poisson point process, which naturally handles three essential features of the problem: (i) the observed sources are a sparse, discrete sampling of a continuous underlying density field; (ii) the detection probability varies with position (the selection function); and (iii) each source's measured position is a noisy realisation of its true position, with a known covariance.  The IPP formalism has become standard in hierarchical Bayesian population studies in astronomy \cite{Mandel2019,Farr2019}, and we follow that framework closely.

The observed catalogue of $N_{\rm det}$ DWDs is modelled as a realisation of an inhomogeneous Poisson point process on the Galactic volume.  The expected number of detections at position $\mathbf{x}$ (in Galactic Cartesian coordinates) is given by the rate density
\begin{equation}
\lambda(\mathbf{x}\mid\Theta)
   = N_{\rm tot}\,
     \rho(\mathbf{x}\mid\Theta)\,
     \bar{\eta}(\mathbf{x}),
\label{eq:rate}
\end{equation}
where $N_{\rm tot}$ is the total number of Galactic DWDs (absorbed into the normalisation; see below) and $\bar{\eta}(\mathbf{x})$ is the spatially-dependent selection function, i.e. the probability that a source at $\mathbf{x}$ is detected, averaged over all other source parameters.  Intuitively, $\lambda(\mathbf{x}\mid\Theta)\,d^{3}x$ is the expected number of detections in a volume element $d^{3}x$ centred at $\mathbf{x}$, and the product $N_{\rm tot}\,\rho(\mathbf{x}\mid\Theta)\,d^{3}x$ gives the number of DWDs physically present there; the selection function $\bar{\eta}(\mathbf{x})$ then determines what fraction of those are actually observed.

For this work we adopt a simple hard-sphere selection function,
\begin{equation}
\bar{\eta}(\mathbf{x}) = \Theta(R_{\rm max} - |\mathbf{x}|),
\qquad R_{\rm max}=15\;\mathrm{kpc},
\label{eq:selection}
\end{equation}
where $\Theta(\cdot)$ is the Heaviside step function.  This choice reflects the fact that the search catalogues used here have negligible completeness beyond $\sim15$~kpc for the observation times considered.  A more realistic $\bar{\eta}(\mathcal{M}_{c},f,R)$ that depends on chirp mass, frequency, and distance (obtained from end-to-end injection-recovery simulations) is deferred to future work; the hard-sphere form is sufficient to prevent the pathological behaviour $R_{d}\to\infty$ that would occur in an uncorrected fit.

The observed catalogue is a realisation of an IPP with rate density $\lambda(\mathbf{x}\mid\Theta)=N_{\rm tot}\,\rho(\mathbf{x}\mid\Theta)\,\bar{\eta}(\mathbf{x})$.  The full Poisson log-likelihood for $N_{\rm det}$ detections at (true) positions $\{\mathbf{y}_i\}$ is
\begin{eqnarray}
\log\mathcal{L}(\Theta,N_{\rm tot})
   = \sum_{i=1}^{N_{\rm det}}
     \log\lambda(\mathbf{y}_i\mid\Theta) \nonumber
     \\ - \int\lambda(\mathbf{y}\mid\Theta)\,d^{3}y.
\label{eq:ipp_poisson}
\end{eqnarray}
The first term is the sum of log-rates at the detected positions; the second is the Poisson normalisation, the expected total number of detections.

Each detected source has a measured position $\mathbf{x}_{i}^{\rm meas}$ that is a noisy observation of its true position $\mathbf{y}$, drawn from a Gaussian with covariance $\mathbf{C}_{\mathbf{x}}^{(i)}$ given by Eq.~\eqref{eq:fisher_prop}.  Marginalising over the unknown true positions, and absorbing $N_{\rm tot}$ into the detection-product term, we have
\begin{multline}
\log\mathcal{L}(\Theta,N_{\rm tot})
   = N_{\rm det}\log N_{\rm tot} \\
     + \sum_{i=1}^{N_{\rm det}}
       \log\!\int \rho(\mathbf{y}\mid\Theta)\,
              \bar{\eta}(\mathbf{y})\,
              \mathcal{N}\bigl(\mathbf{x}_{i}^{\rm meas}\mid
                                 \mathbf{y},\,
                                 \mathbf{C}_{\mathbf{x}}^{(i)}\bigr)\,
              d^{3}y \\
     - N_{\rm tot}\int \rho(\mathbf{y}\mid\Theta)\,
         \bar{\eta}(\mathbf{y})\,d^{3}y,
\label{eq:ipp_full}
\end{multline}
where $\mathcal{N}(\mathbf{x}_{i}^{\rm meas}\mid\mathbf{y},\mathbf{C}_{\mathbf{x}}^{(i)})$ denotes the Gaussian position-uncertainty kernel for source $i$.

The total number of Galactic DWDs $N_{\rm tot}$ is unknown and is not of direct interest for structure inference.  We eliminate it by profiling, i.e. maximising $\log\mathcal{L}$ with respect to $N_{\rm tot}$ for each $\Theta$.  Setting $\partial(\log\mathcal{L})/\partial N_{\rm tot}=0$ gives the conditional maximum-likelihood estimate
\begin{equation}
\hat{N}_{\rm tot}(\Theta)
   = \frac{N_{\rm det}}
          {\int\rho(\mathbf{y}\mid\Theta)\,\bar{\eta}(\mathbf{y})\,d^{3}y}
   \equiv \frac{N_{\rm det}}{Q(\Theta)},
\label{eq:ntot_hat}
\end{equation}
where $Q(\Theta)\equiv\int\rho\,\bar{\eta}\,d^{3}y$ is the detectable-mass fraction, i.e. the probability that a randomly drawn DWD falls within the selection volume.  Substituting $\hat{N}_{\rm tot}$ back and discarding additive constants yields the working log-likelihood
\begin{multline}
\log\mathcal{L}(\Theta)
   = \sum_{i=1}^{N_{\rm det}}
     \log\!\int \rho(\mathbf{y}\mid\Theta)\,
            \bar{\eta}(\mathbf{y})\,
            \mathcal{N}_{i}(\mathbf{y})\,d^{3}y \\
     \;-\; N_{\rm det}\log Q(\Theta).
\label{eq:ipp_logL}
\end{multline}

Equation~\eqref{eq:ipp_logL} is the central likelihood of \textsc{Galena}.  The first term evaluates the density model, convolved with each source's measurement uncertainty, at the observed positions.  The second term penalises models that spread the detectable mass too broadly: a model with large $R_d$ or $z_d$ would cover the entire Galaxy, making $Q(\Theta)$ large and reducing the per-source probability density through the $-N_{\rm det}\log Q$ term.

For truth catalogues (LDC) and synthetic data, the true source positions are known exactly.  In this limit the Gaussian kernel $\mathcal{N}_i(\mathbf{y})$ reduces to a Dirac delta, and Eq.~\eqref{eq:ipp_logL} simplifies to
\begin{multline}
\log\mathcal{L}(\Theta)
   = \sum_{i=1}^{N_{\rm det}} \log\rho(\mathbf{x}_i^{\rm true}\mid\Theta) \\
     \;-\; N_{\rm det}\log Q(\Theta),
\label{eq:ipp_exact}
\end{multline}
which requires no convolution and no pre-computed $\mathbf{W}$ matrix.  \textsc{Galena} automatically selects this fast exact path when the source type is not a search pipeline output (Sec.~\ref{sec:data}).  For GBSIEVER search results, where the reported positions carry real measurement errors, the full convolution path via Eq.~\eqref{eq:ipp_logL} is used.

The priors on $\Theta$ are configurable through the inference configuration file, which supports uniform, truncated-Gaussian, and log-normal forms independently for each parameter.  All results reported in this paper adopt uniform priors on all four parameters, over the ranges $A\in[0,1]$, $R_b\in[100,5000]$~pc, $R_d\in[500,15000]$~pc, and $z_d\in[50,3000]$~pc.  These ranges bracket the literature values of the Milky Way \cite{BlandHawthorn2016,Bovy2012,Juric2008} while remaining wide enough that the posterior is driven entirely by the data; in particular, none of the reported posteriors touches a prior boundary (Sec.~\ref{sec:results}).

The posterior $p(\Theta\mid\mathcal{D})\propto\mathcal{L}(\Theta)\,\pi(\Theta)$ is sampled with \texttt{dynesty}~\cite{Speagle2020}, a nested-sampling package that jointly estimates the Bayesian evidence and the posterior.  Nested sampling is well suited to this problem: it requires no proposal tuning, is robust to the mild degeneracies of the four-dimensional posterior, and its evidence estimate $\log Z$ enables the model-comparison studies discussed in Sec.~\ref{sec:conclusions}.  All runs use static nested sampling with $N_{\rm live}=1000$ live points, a stopping criterion of $\Delta\ln Z<0.1$, and the default multi-ellipsoidal bound with random-walk proposals, with the unit-cube-to-parameter prior transform constructed analytically from the priors above.  A typical run requires $\sim10^{4}$--$10^{5}$ likelihood evaluations, which the sparse $\mathbf{W}$-matrix of Sec.~\ref{sec:sparse_W} makes computationally inexpensive.

\subsection{Pre-computed sparse convolution matrix}
\label{sec:sparse_W}

The log-likelihood in Eq.~\eqref{eq:ipp_logL} presents a computational obstacle: the inner integral is a three-dimensional convolution of the density model $\rho(\mathbf{y}\mid\Theta)$ with $N_{\rm det}$ Gaussian kernels, each with its own $3\times3$ covariance $\mathbf{C}_{\mathbf{x}}^{(i)}$.  Evaluating this integral numerically at every likelihood call would be prohibitively expensive: during nested sampling the likelihood is evaluated $\sim10^{4}$--$10^{5}$ times, and a naive per-source-per-grid-point quadrature would require $\mathcal{O}(N_{\rm det}\times N_{\rm grid})\sim10^{6}$--$10^{8}$ density evaluations per likelihood call.  The Gaussian measurement kernel is independent of $\Theta$, which allows the convolution to be performed once, before sampling; each likelihood evaluation then reduces to a fast matrix--vector product.

We discretise the Galactic volume onto a uniform Cartesian grid $\{\mathbf{y}_{j}\}_{j=1}^{N_{\rm grid}}$ with cell volume $\Delta V = \Delta x\,\Delta y\,\Delta z$.  Defining the pre-computed \emph{weight matrix} $\mathbf{W}\in\mathbb{R}^{N_{\rm grid}\times N_{\rm det}}$ via
\begin{equation}
W_{ji}
   \equiv \bar{\eta}(\mathbf{y}_{j})\,
          \mathcal{N}\bigl(\mathbf{y}_{j}\mid
                            \mathbf{x}_{i}^{\rm meas},\,
                            \mathbf{C}_{\mathbf{x}}^{(i)}\bigr)\,
          \Delta V,
\label{eq:W_def}
\end{equation}
the detection-term integral for source $i$ becomes a discrete sum
\begin{multline}
\int \rho(\mathbf{y}\mid\Theta)\,\bar{\eta}(\mathbf{y})\,
      \mathcal{N}\bigl(\mathbf{x}_{i}^{\rm meas}\mid
                         \mathbf{y},\,
                         \mathbf{C}_{\mathbf{x}}^{(i)}\bigr)\,d^{3}y \\[2pt]
   \approx \sum_{j=1}^{N_{\rm grid}}
               \rho_{j}(\Theta)\,W_{ji},
\label{eq:conv_discrete}
\end{multline}
where $\rho_{j}(\Theta)\equiv\rho(\mathbf{y}_{j}\mid\Theta)$.  The log-likelihood~\eqref{eq:ipp_logL} then simplifies to
\begin{multline}
\log\mathcal{L}(\Theta)
   = \sum_{i=1}^{N_{\rm det}}
     \log\!\Bigl(\sum_{j} \rho_{j}(\Theta)\,W_{ji}\Bigr) \\
     - N_{\rm det}\,
       \log\!\Bigl(\sum_{j}\rho_{j}(\Theta)\,\bar{\eta}_{j}\,\Delta V\Bigr),
\label{eq:ipp_fast}
\end{multline}
where $\bar{\eta}_{j}\equiv\bar{\eta}(\mathbf{y}_{j})$.  The key advantage of this formulation is that all $\Theta$-dependence is isolated in the density vector $\bm{\rho}(\Theta)$; the matrix $\mathbf{W}$ is computed once and reused.

The matrix $\mathbf{W}$ is constructed once, before sampling begins, and stored in compressed sparse row format.  Sparsity is achieved by truncating each source's Gaussian kernel at a local radius $r_{\rm local}\simeq 4\;\mathrm{kpc}$, beyond which the kernel weight is negligible.  This yields a sparsity of order $1\%$ at the grid resolutions used here.  Each likelihood evaluation in Eq.~\eqref{eq:ipp_fast} then requires only two sparse matrix--vector products, $\bm{\rho}^{\mathsf{T}}\mathbf{W}$ for the detection term and $\bm{\rho}^{\mathsf{T}}\bar{\bm{\eta}}$ for the normalisation, achieving $\sim700$ likelihood evaluations per second at $21^{3}$, $\sim40$ at $51^{3}$, and $\sim5$ at the $101^{3}$ resolution of the main results, on a single CPU core.  This speed makes nested sampling with \texttt{dynesty} \cite{Speagle2020} practical even for catalogues of several thousand sources.

The grid resolution represents a trade-off between accuracy and speed.  For the search-pipeline results of Sec.~\ref{sec:results} we use $101^{3}=1\,030\,301$ cells ($\Delta x=\Delta y\approx 297\;\mathrm{pc}$, $\Delta z\approx 79\;\mathrm{pc}$), placing $\sim4$ grid cells across the disk scale height $z_{d}\sim 300\;\mathrm{pc}$; truth-catalogue runs use $51^{3}$ cells, and quick-validation runs $21^{3}$--$31^{3}$.  The resolution requirements are quantified in Sec.~\ref{sec:results}.  The same grid is used for the normalisation integral $Q(\Theta)$ in Eq.~\eqref{eq:ntot_hat}.

Together with the analytical Jacobian of Sec.~\ref{sec:jacobian}, this sparse pre-computation strategy enables \textsc{Galena} to perform full Bayesian inference on catalogues of thousands of DWDs in hours rather than days. The pipeline operates in two modes: a synthetic mode, where sources are drawn from the density model with known truth for validation, and a truth-catalogue mode, which ingests catalogues of measured DWD parameters as described below.

\section{Data}
\label{sec:data}

We use several complementary DWD catalogues, spanning both truth-level (injection) and search-level (detected) data, across two millihertz detectors.  Sec.~\ref{subsec:dwd_cata} describes the catalogues, and Sec.~\ref{subsec:source_select} defines the quality cuts applied to them together with the per-source Fisher-matrix position covariances computed for the retained sources, whose SNR dependence shapes the error model entering the likelihood.

\subsection{DWD catalogues}\label{subsec:dwd_cata}

These serve complementary roles: the injection catalogues, being free of measurement error, provide the reference against which the pipeline is validated, while the search catalogues carry the realistic measurement errors and selection effects that the hierarchical model must propagate.

For truth-level data we use the LISA Data Challenge catalogue LDC1-4 GB v2 \cite{LDC2021guide,babak2020lisa,babak2010mock,baghi2022lisa}, the standardised Galactic DWD catalogue for LISA, in which each source carries eight true parameters together with the SNR for a single LISA detector.

For search-level data we apply the GBSIEVER iterative-fit pipeline \cite{zhang2021pso} to the LDC LISA data stream, built on matched-filtering searches \cite{Jaranowski1998} and time-delay-interferometry data preprocessing \cite{tinto2014time}.  GBSIEVER resolves sources iteratively: in each frequency band it maximises the $\mathcal{F}$-statistic over the intrinsic parameters with particle-swarm optimisation (Sec.~\ref{sec:signal}), subtracts the best-fit signal from the data, and repeats until a termination criterion is met \cite{zhang2021pso}.  On the LDC1-4 data the plain search identifies $32{,}987$ candidate sources, of which $12{,}251$ survive cross-validation against a secondary search and $10{,}388$ are confirmed against the injected catalogue \cite{zhang2021pso}.  Following the terminology of \cite{zhang2021pso,universe11080248}, the cross-validation survivors are the \emph{reported} sources---the primary-search detections that pass the frequency- and SNR-dependent cross-correlation threshold---while \emph{confirmed} sources are the subset of reported sources that also match an entry of the injected (true) catalogue.  Throughout this paper ``GBSIEVER-reported'' denotes this reported (blind-search) class, as distinct from the injected truth catalogue.  The reported output per source is the maximum-likelihood estimate of the eight waveform parameters $\bm{\theta}$ together with an SNR; no per-source covariance is provided, which motivates computing the per-source Fisher matrix directly from the waveform model (Sec.~\ref{sec:fim}).  The same pipeline, in its prior-off variant, is also applied to the Taiji data stream in TDC I.

Unlike the truth catalogue, these are \emph{measured} parameters with realistic search-level uncertainties and detection biases, and they constitute the primary science dataset of this work.  We analyse two variants of the LDC search: \emph{prior off}, which performs the standard iterative fit without additional constraints, and \emph{prior on}, which augments the per-source parameter estimation with an astrophysically motivated prior on the DWD frequency derivative. This prior encodes the expectation that the derivative lies within a specified range that scales with the frequency value \cite{Zhao:2026dse}.  The prior regularises the $\dot{f}$ (and hence distance) estimates of low-SNR sources toward physically plausible values; after the quality cuts of Sec.~\ref{sec:results}, however, the retained source lists are nearly identical with and without the prior, so its impact on the final inference is negligible.

\subsection{Source selection and parameters}\label{subsec:source_select}

All catalogues provide the five GW parameters $\bm{\theta}=(f,\dot{f},\beta,\lambda,\mathcal{A})$ and a per-source SNR; for the truth catalogues these are the injected values, while for the search catalogues they are the GBSIEVER maximum-likelihood estimates.  From the full GBSIEVER LDC catalogue of $\sim10^{4}$ detections we retain the sources that pass a sequence of quality cuts: $\dot{f}>1/T_{\rm obs}^{2}$ with $T_{\rm obs}=2$~yr (the theoretical frequency resolution over the observation time), $\mathrm{SNR}>10$, a Galactocentric radius $|\mathbf{x}_{\rm Gal}|<R_{\rm max}=15\;\mathrm{kpc}$ (the selection boundary of Eq.~\eqref{eq:selection}), and a nearest-neighbour cut in Galactic position $\mathrm{NN}<500$~pc, which discards isolated sources whose poorly constrained distances place them far from the clustered disk/bulge population.  These cuts retain $N_{\rm det}=2175$ sources (LDC on), $N_{\rm det}=2151$ (LDC off), and $N_{\rm det}=2170$ (TDC off); the main runs use the full retained samples without sub-sampling, while earlier quick-validation runs drew $N=2000$--$5000$ sources for speed.  Detected sources span SNR $\sim5$--$1200$ with the majority below $\sim50$, and the main runs retain those above SNR 10, as used in Sec.~\ref{sec:results}.

Because none of the catalogues provide per-source Fisher matrices, we compute each source's $6\times6$ waveform Fisher information matrix and marginalise it to $\bm{\theta}=(f,\dot{f},\beta,\lambda,\mathcal{A})$ as described in Sec.~\ref{sec:fim}, then propagate it to position space via Eq.~\eqref{eq:fisher_prop}.

The SNR distribution matters for the error model.  Because the Fisher position error scales inversely with SNR, $\sigma_{R}/R\propto\mathrm{SNR}^{-1}$, the low-SNR majority of the catalogue contributes the broadest position kernels to the $\mathbf{W}$ matrix.  The $\mathrm{SNR}>10$ cut therefore trades catalogue size against the dominance of poorly localised sources whose kernels would smear the thin-disk structure beyond recovery, while the $\mathrm{NN}<500$~pc cut serves the complementary role of removing sources whose distances are so uncertain that they are effectively unlocalised with respect to the clustered disk and bulge.  The retained sources span SNR $\sim10$--$1200$, with fractional distance uncertainties of a few tens of percent at the $\mathrm{SNR}\sim10$ threshold, falling as $\mathrm{SNR}^{-1}$ to a few tenths of a percent for the loudest sources.

\section{Results}
\label{sec:results}

We present the results in five parts.  We first validate the pipeline end-to-end on synthetic injection-recovery data (Sec.~\ref{sec:synth_results}), which also establishes the grid-resolution requirements for unbiased recovery; we then apply it to the GBSIEVER-reported LDC and TDC catalogues (Sec.~\ref{subsec:LDCTDC}), cross-checking the recovered posterior against an independent particle-swarm optimisation and characterising the per-source position uncertainties; we compare these search-level results with the LDC truth catalogue to isolate the impact of measurement errors (Sec.~\ref{subsec:LDCTruth}); we quantify the effect of the search-stage astrophysical prior (Sec.~\ref{subsec:LDCprior}); and we close by demonstrating that the $101^3$ grid used throughout is already converged (Sec.~\ref{subsec:grid}).

\subsection{Validation on synthetic data}
\label{sec:synth_results}

We first validate the pipeline end-to-end with synthetic injection-recovery tests.  Sources are drawn from the two-component density model at the canonical literature parameters $\Theta=(0.25,\,500,\,2500,\,300)$ (units of pc for the three scale parameters; these differ from the LDC truth catalogue's own fitted values), generated by rejection sampling $N_0=3000$ candidates; the measured catalogue is obtained by applying the hard selection ($R<15$~kpc) and perturbing the surviving positions with diagonal Gaussian noise $\sigma_R=\max(0.15R,\,50\;\mathrm{pc})$, mimicking per-source measurement error.  The full $\mathbf{W}$-matrix likelihood path of Eq.~\eqref{eq:ipp_fast} is then used to recover the posterior, with uniform priors, $N_{\rm live}=250$, and a stopping tolerance $\Delta\ln Z<0.5$, on $21^3$, $31^3$, and $51^3$ integration grids to assess grid-resolution convergence.

\begin{table*}
\caption{Synthetic injection-recovery results as a function of grid resolution (truth $\Theta=(0.25,500,2500,300)$; $N_0=3000$ sources, uniform priors, $N_{\rm live}=250$, $\Delta\ln Z<0.5$).  Cells give the posterior median with 68\% credible interval.  The $21^3$ grid under-resolves the bulge ($Q\simeq1.40>1$) and biases $A$ low; the artifact largely disappears at $31^3$ ($Q\simeq1.07$) and is negligible at $51^3$ ($Q\simeq0.99$), where all four parameters are recovered close to their injected values.}
\label{tab:synth_results}
\begin{center}
\renewcommand{\arraystretch}{1.4}
\small\setlength{\tabcolsep}{4pt}
\begin{tabular}{lcccc}\toprule
Param  & Truth & $21^3$ & $31^3$ & $51^3$ \\\midrule
$A$           & 0.25 & $0.10\,[0.10,0.11]$ & $0.21\,[0.20,0.23]$ & $0.25\,[0.24,0.26]$ \\
$R_b$ (pc)    & 500  & $509\,[492,526]$     & $582\,[570,594]$     & $530\,[520,541]$ \\
$R_d$ (pc)    & 2500 & $2344\,[2303,2386]$  & $2358\,[2310,2406]$  & $2407\,[2362,2455]$ \\
$z_d$ (pc)    & 300  & $331\,[317,345]$     & $305\,[289,320]$     & $296\,[277,312]$ \\
\bottomrule
\end{tabular}
\end{center}
\end{table*}

Table~\ref{tab:synth_results} summarises the recovered parameters as a function of grid resolution.  The disk parameters are recovered accurately even on the coarsest ($21^3$) grid: $R_d=2344$~pc (truth $2500$) and $z_d=331$~pc (truth $300$).  The bulge fraction, however, is biased low ($A=0.10$ versus the truth $0.25$).  This is a grid-resolution artifact: at $21^3$ resolution ($\Delta x\approx1430$~pc $\gg R_b=500$~pc) the midpoint rule over-integrates the bulge, producing a detectable-mass fraction $Q=\int\rho\,\eta\,dV\simeq1.40>1$, and the $-N_{\rm det}\log Q$ term in Eq.~\eqref{eq:ipp_logL} over-penalises the bulge fraction.  At $31^3$ ($Q\simeq1.07$) the artifact largely disappears and the recovered $A$ rises to $\simeq0.21$; at $51^3$ ($Q\simeq0.99$) the bias is negligible for $A$ and $z_d$ ($A=0.25$, $z_d=296$~pc, both within their credible intervals), while $R_b$ and $R_d$ remain mildly biased ($R_b=530$~pc and $R_d=2407$~pc versus the injected $500$ and $2500$~pc), reflecting the residual bulge--disk degeneracy.  These tests confirm that the IPP likelihood with sparse $\mathbf{W}$ pre-computation correctly recovers the disk structural parameters from a ``noisy'', selection-biased catalogue, and they motivate the grid-resolution requirements verified against the real search data in Sec.~\ref{subsec:grid}.

\subsection{GBSIEVER search catalogue: \texorpdfstring{$101^3$}{101³} results}
\label{subsec:LDCTDC}
We apply \textsc{Galena} to GBSIEVER-reported LDC and TDC sources, using the full $\mathbf{W}$-matrix likelihood path to propagate per-source measurement uncertainties.  Sources are selected with SNR$>10$, $\dot{f}>1/T_{\rm obs}^{2}$ ($T_{\rm obs}=2$~yr), Galactocentric radius $R<15$~kpc, and nearest-neighbour distance NN$<500$~pc (to discard isolated outliers with poorly constrained distances).  These cuts retain $N_{\rm det}=2175$ sources (LDC on), $N_{\rm det}=2151$ (LDC off), and $N_{\rm det}=2170$ (TDC off).  All runs use $N_{\rm live}=1000$, a stopping tolerance $\Delta\ln Z<0.1$, uniform priors, and a $101^3$ integration grid ($\Delta x \approx 297$~pc, $\Delta z \approx 79$~pc), whose convergence is verified in Sec.~\ref{subsec:grid}.

\begin{table*}
\caption{Posterior medians and 68$\%$ credible intervals for GBSIEVER-reported sources, all on $101^3$ grids ($N_{\rm live}=1000$, uniform priors).  LDC and TDC are single detector runs \protect\cite{zhang2021pso,Zhang2025vya}; the LDC (LISA+Taiji-like) column is the joint two-detector run\cite{zhang2022resolving}, in which a detector with the LISA arm length ($2.5$~Gm) is placed at the Taiji orbit position.  Literature values from \protect\cite{BlandHawthorn2016} are shown for reference.}
\label{tab:gbsiever_results}
\begin{center}
\renewcommand{\arraystretch}{1.4}
\small\setlength{\tabcolsep}{4pt}
\begin{tabular}{lcccc}\toprule
Param  & LDC (LISA) & TDC (Taiji) & LDC (LISA+Taiji-like) & Literature \\\midrule
$A$           & $0.187^{+0.011}_{-0.012}$ & $0.222^{+0.012}_{-0.012}$ & $0.204^{+0.011}_{-0.011}$ & 0.25 \\
$R_b$ (pc)    & $773^{+26}_{-25}$  & $700^{+19}_{-19}$  & $749^{+22}_{-21}$  & 500 \\
$R_d$ (pc)    & $2175^{+51}_{-52}$ & $2234^{+57}_{-53}$ & $2210^{+54}_{-53}$ & 2500 \\
$z_d$ (pc)    & $282^{+7}_{-7}$  & $276^{+7}_{-7}$  & $281^{+7}_{-7}$  & 300 \\\bottomrule
\end{tabular}
\end{center}
\end{table*}

Table~\ref{tab:gbsiever_results} summarises the posterior constraints.  The LDC on and LDC off results are nearly identical, indicating that the search-stage astrophysical prior affects the composition of the pre-cut catalogue rather than the final source list or the posterior.  The TDC catalogue yields systematically lower $R_b$ and $z_d$ and a higher $A$, likely reflecting the different detector configuration adopted in the TDC simulation (the longer Taiji arm length). The joint LISA+Taiji-like run on the LDC catalogue yields $R_b=749$~pc, $R_d=2210$~pc, $z_d=281$~pc and $A=0.204$, a systematic shift relative to the LISA-only result; the credible intervals shrink for $R_b$ but not for $R_d$ or $z_d$ (Table~\ref{tab:gbsiever_results}), as discussed in Sec.~\ref{sec:conclusions}.  The value $R_b\approx773$~pc is elevated relative to the canonical stellar-mass value ($\sim500$~pc), whereas $z_d\approx282$~pc is consistent with the canonical thin-disk scale height ($\sim300$~pc); $R_b$ remains the most weakly constrained of the scale parameters and should be interpreted with caution.

As an independent cross-check, we also maximise the same IPP likelihood with particle-swarm optimisation (PSO) instead of nested sampling.  The PSO best-fit parameters agree with the dynesty posterior medians to within a few tenths of a percent for every parameter in all four runs (e.g.\ $R_b=772$ versus $773$~pc and $z_d=282$ versus $282$~pc for the LDC prior-off run), confirming that the posteriors are well converged and not an artefact of the sampler.  The improvement of the present results over our earlier constraints---in particular the consistency of $z_d$ with the canonical $\sim300$~pc and the larger recovered bulge fraction $A$---is most plausibly attributed to the likelihood itself: the per-source position covariances are now built from the waveform Fisher matrix (Sec.~\ref{sec:fim}) rather than neglecting the estimation error.  The agreement of the two maximisers establishes that the posteriors are converged and rules out the sampler as the source of the shift, but does not by itself prove that the likelihood change is the cause; that attribution rests on the likelihood being the only material change between the two analyses.  Table~\ref{tab:pso_dynesty} quantifies this agreement.

\begin{table*}
\caption{Cross-check of the particle-swarm optimisation (PSO) best fit against the \texttt{dynesty} posterior median for the four search-catalogue runs.  Each cell lists PSO / \texttt{dynesty}.  The two maximisers agree to within a fraction of a percent for every parameter.}
\label{tab:pso_dynesty}
\begin{center}
\renewcommand{\arraystretch}{1.4}
\small\setlength{\tabcolsep}{6pt}
\begin{tabular}{lcccc}\toprule
Run & $A$ & $R_b$ (pc) & $R_d$ (pc) & $z_d$ (pc) \\\midrule
LDC (LISA)       & $0.187$/$0.187$ & $772$/$773$ & $2171$/$2175$ & $282$/$282$ \\
LDC on (LISA)    & $0.190$/$0.190$ & $767$/$769$ & $2179$/$2181$ & $281$/$282$ \\
TDC (Taiji)      & $0.222$/$0.222$ & $699$/$700$ & $2233$/$2234$ & $276$/$276$ \\
LDC (LISA+Taiji-like) & $0.204$/$0.204$ & $748$/$749$ & $2209$/$2210$ & $281$/$281$ \\\bottomrule
\end{tabular}
\end{center}
\end{table*}

Several features of the posterior deserve comment.  First, the formal uncertainties are a few percent ($\sim3\%$ for $R_b$, $\sim2\%$ for $R_d$, $\sim3\%$ for $z_d$, and $\sim6\%$ for $A$).  These are \emph{statistical} uncertainties from the IPP likelihood and do not account for the dominant systematic: the mapping from the two-component parametric model to the true DWD spatial distribution.  The recovered $z_d\approx282$~pc is consistent with the canonical thin-disk value of $\sim300$~pc \cite{reid2014trigonometric,bennett2019vertical}, and the recovered $A=0.187$ lies closer to the literature value of $0.25$ than in our earlier exact-position analysis~\cite{universe11080248}.  The offset of $R_b$ ($\sim270$~pc above the canonical $500$~pc) is harder to attribute to astrophysics alone: $R_b$ is weakly constrained and the bulge--disk degeneracy is only partially resolved, so part of this offset may be a systematic consequence of the crude spherical-Gaussian bulge model rather than a genuine radial broadening of the bulge.

Second, the log-evidence values are reported for completeness: $\log Z = -80074$ (LDC on), $-79469$ (LDC off), $-84689$ (TDC off), and $-91539$ (LDC, LISA+Taiji-like).  Because each run conditions on a different source catalogue, these values are not directly comparable and cannot serve as Bayes factors for model comparison; the large offset between the LDC and TDC runs reflects the different source lists, measurement-error patterns, and selection of the two search outputs rather than a difference in fit quality.  Within the LDC LISA runs, the modest $|\Delta\log Z|\simeq605$ between the on and off catalogues likewise reflects the slightly different sample compositions produced by the prior-enhanced search, not a preference between models.

\begin{figure}
\centering
\includegraphics[width=0.48\textwidth]{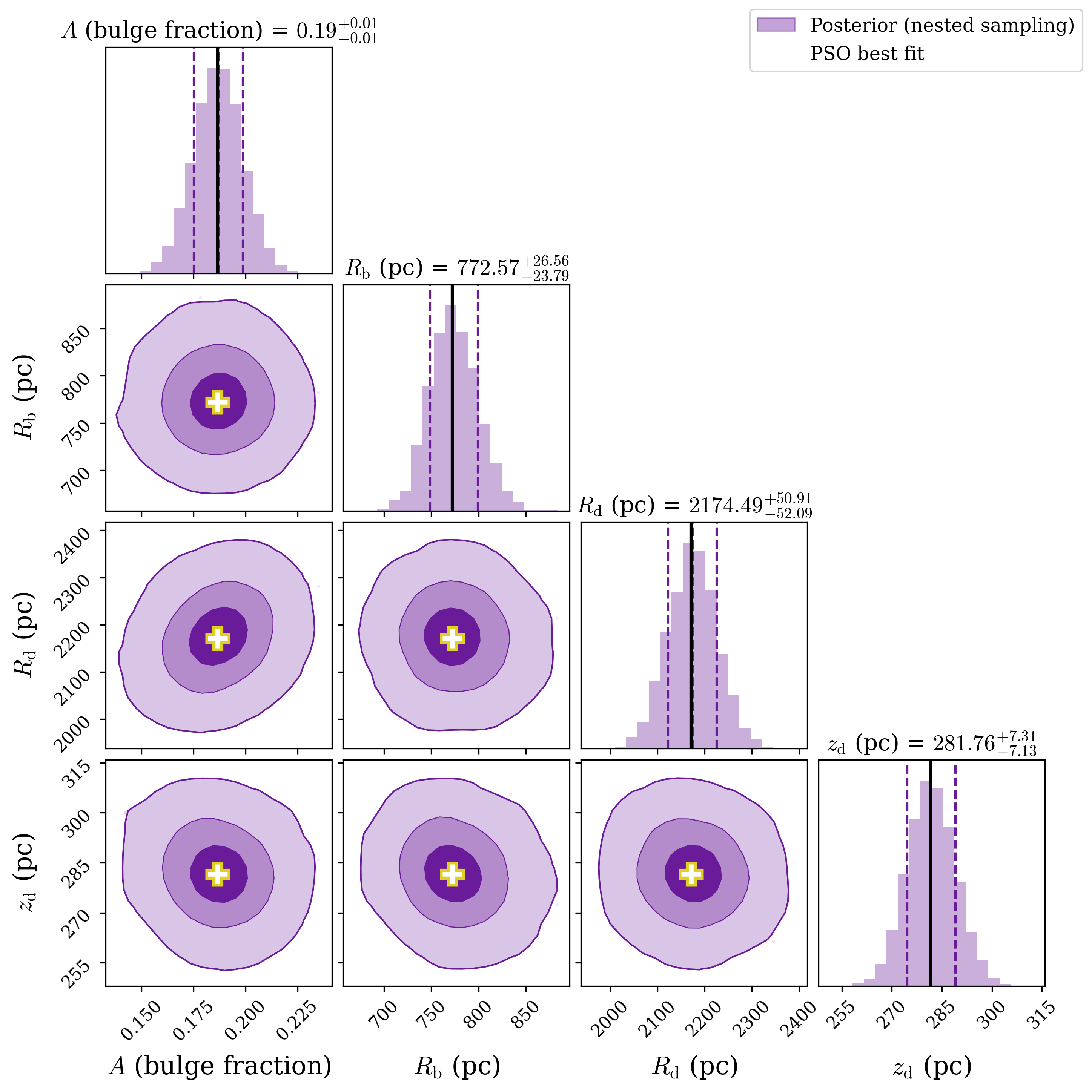}
\caption{Corner plots for GBSIEVER-reported LDC sources ($101^3$ grid). Contours enclose 68$\%$ and 95$\%$ of the resampled posterior mass; the small cross in each two-dimensional panel marks the particle-swarm best fit.}
\label{fig:corner_ldc}
\end{figure}

\begin{figure}
\centering
\includegraphics[width=0.48\textwidth]{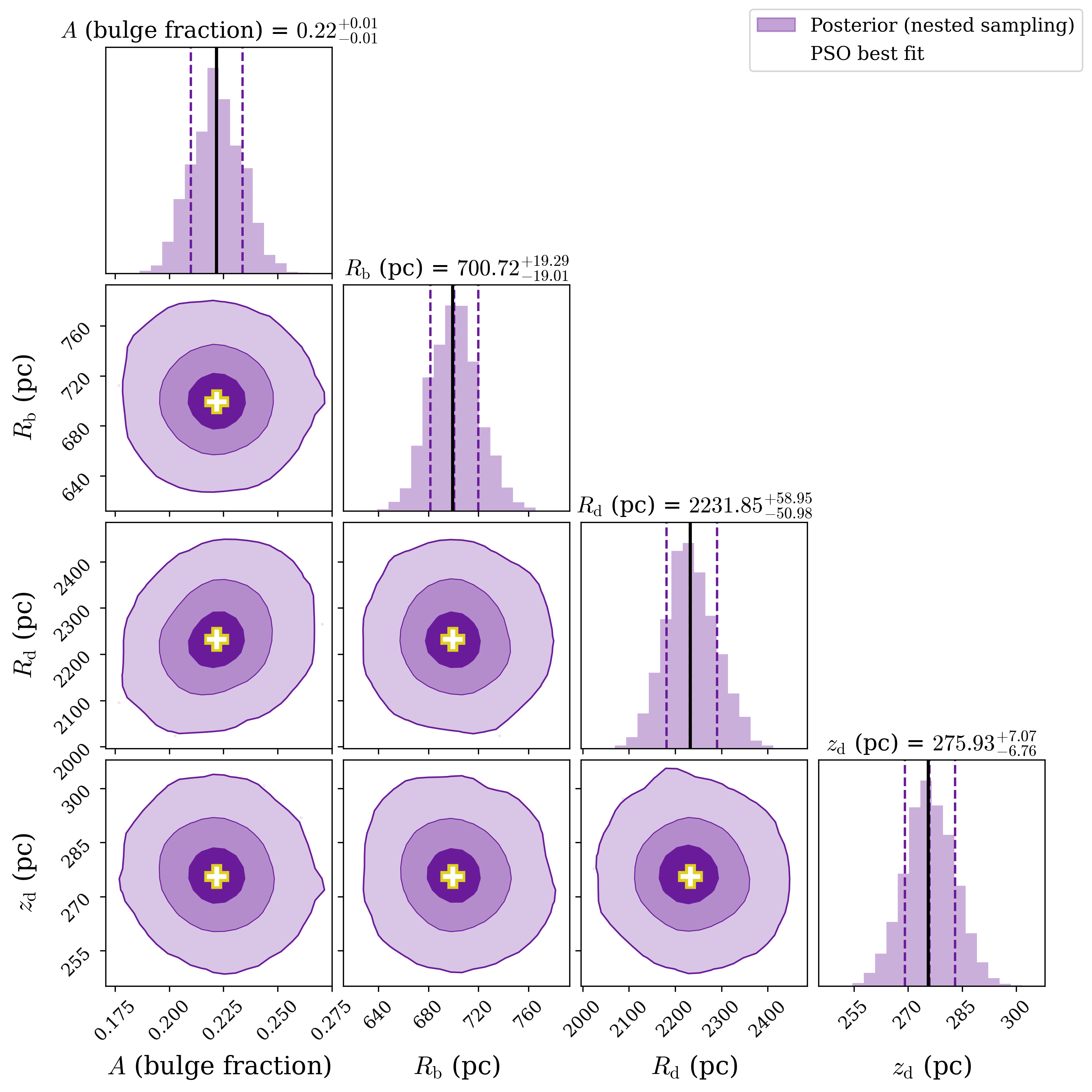}
\caption{Same as Fig.~\ref{fig:corner_ldc} for GBSIEVER-reported TDC sources.}
\label{fig:corner_tdc}
\end{figure}

The corner plots are shown in Figs.~\ref{fig:corner_ldc} and \ref{fig:corner_tdc}.  In both, the one-dimensional marginal posteriors along the diagonal are smooth and unimodal, with no secondary modes or prior-boundary pile-up.  The off-diagonal panels show only weak residual correlations: the $A$--$R_d$ panel is slightly elongated along the positive diagonal, while the $A$--$R_b$, $R_b$--$R_d$ and $R_d$--$z_d$ panels are comparatively round.  The bulge fraction $A$ is correspondingly the least tightly constrained of the four parameters.  In the two-dimensional panels the small cross marks the particle-swarm best fit to the same IPP likelihood; its near-coincidence with the posterior peak reflects the agreement between the particle-swarm and nested-sampling maximisers quantified in Table~\ref{tab:pso_dynesty}.

The two figures are qualitatively similar but differ in detail in a way that tracks the posterior values in Table~\ref{tab:gbsiever_results}.  Relative to the LDC result (Fig.~\ref{fig:corner_ldc}), the TDC corner plot (Fig.~\ref{fig:corner_tdc}) has a tighter $R_b$ marginal ($\pm19$~pc versus $\pm26$~pc) and a comparable $z_d$ marginal ($\pm7$~pc versus $\pm7$~pc), while its $A$ marginal is shifted to larger values ($A=0.222$ versus $0.187$) and its $R_d$ marginal to $\sim2234$~pc.  These shifts mirror the systematic differences between the LDC and TDC source lists reported above, rather than a change in the underlying model. Since the LISA and Taiji detectors have different arm lengths, the two catalogues have different SNR distributions and different per-source position uncertainties, which propagate through the $\mathbf{W}$ matrix to yield the observed shifts in the posterior.

\begin{figure}
\centering
\includegraphics[width=0.48\textwidth]{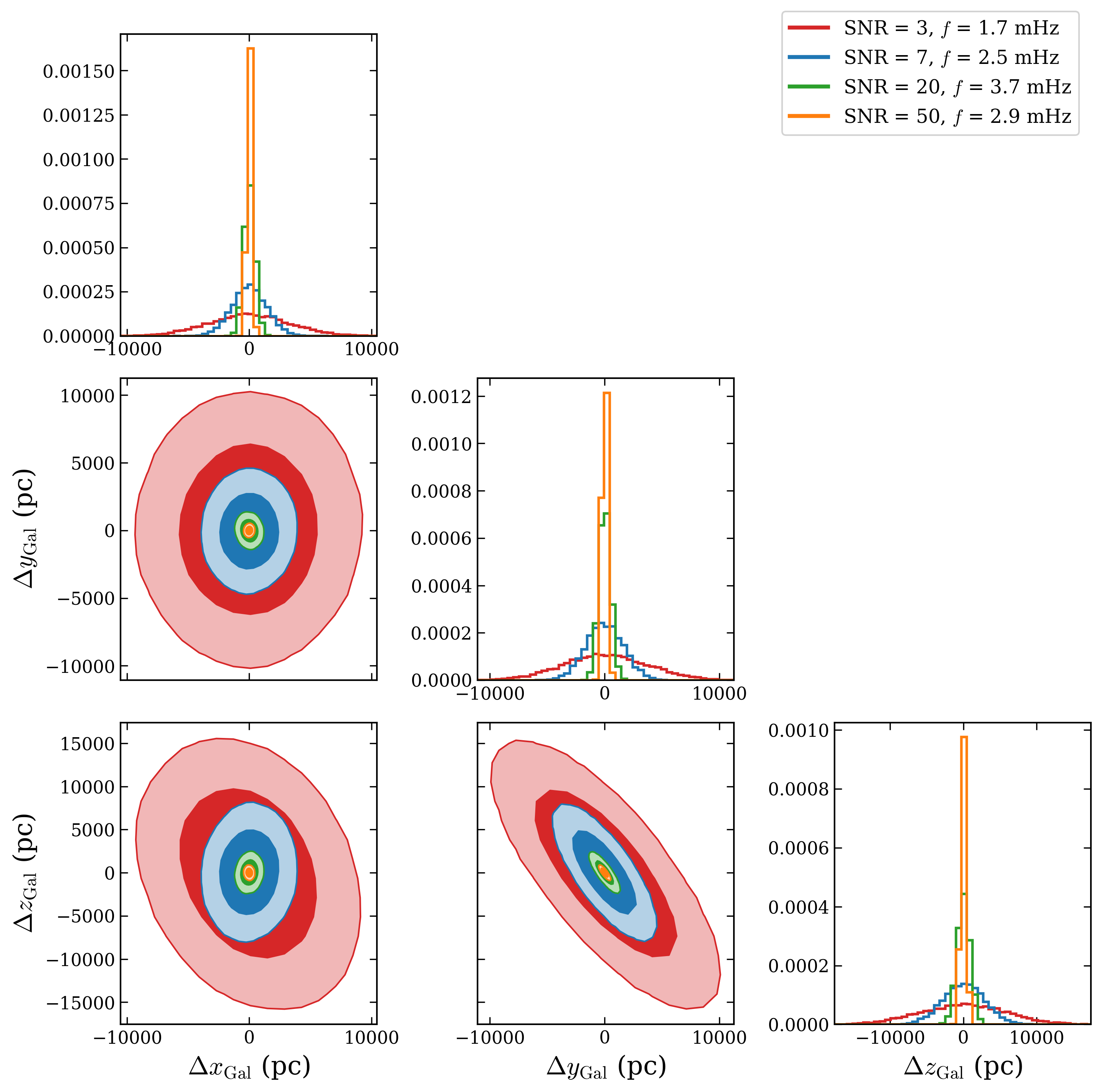}
\caption{Position error distributions for four representative DWD sources with SNR $=3$, 7, 20 and 50 ($f=1.7$, 2.5, 3.7 and 2.9~mHz).  Samples are drawn from the Gaussian $\mathcal{N}(\mathbf{0},\mathbf{C}_{\mathbf{x}}^{(i)})$ of Eq.~\eqref{eq:fisher_prop}; contours enclose 68\% and 95\% of the sampled positions, and the diagonal panels show the one-dimensional marginals.}
\label{fig:snr_unc}
\end{figure}

To make the per-source errors entering the $\mathbf{W}$ matrix concrete, Fig.~\ref{fig:snr_unc} compares the position error ellipsoids of four representative DWD sources spanning SNR $=3$, 7, 20 and 50 ($f=1.7$, 2.5, 3.7 and 2.9~mHz).  Each panel samples the Gaussian $\mathcal{N}(\mathbf{0},\mathbf{C}_{\mathbf{x}}^{(i)})$ of Eq.~\eqref{eq:fisher_prop}, constructed from the waveform Fisher matrix of Sec.~\ref{sec:fim}.  The total position uncertainty $\delta R=\sqrt{\mathrm{Tr}\,\mathbf{C}_{\mathbf{x}}^{(i)}}$ spans $7679$~pc (SNR 3), $3592$~pc (SNR 7), $1110$~pc (SNR 20) and $419$~pc (SNR 50), following the expected $\mathrm{SNR}^{-1}$ scaling to within $\sim10\%$; the residual deviations reflect the different frequencies and sky positions of the four sources.  These ellipsoids are exactly the Gaussian kernels convolved into the $\mathbf{W}$ matrix in Eq.~\eqref{eq:W_def}: a low-SNR source smears its likelihood weight over a kiloparsec-scale volume, far larger than the $\Delta x\approx300$~pc grid spacing, so that the likelihood responds to the smoothed source density rather than to the point sources themselves.

\subsection{Comparison with the truth catalogue}\label{subsec:LDCTruth}

\begin{figure}[t]
\centering
\includegraphics[width=0.48\textwidth]{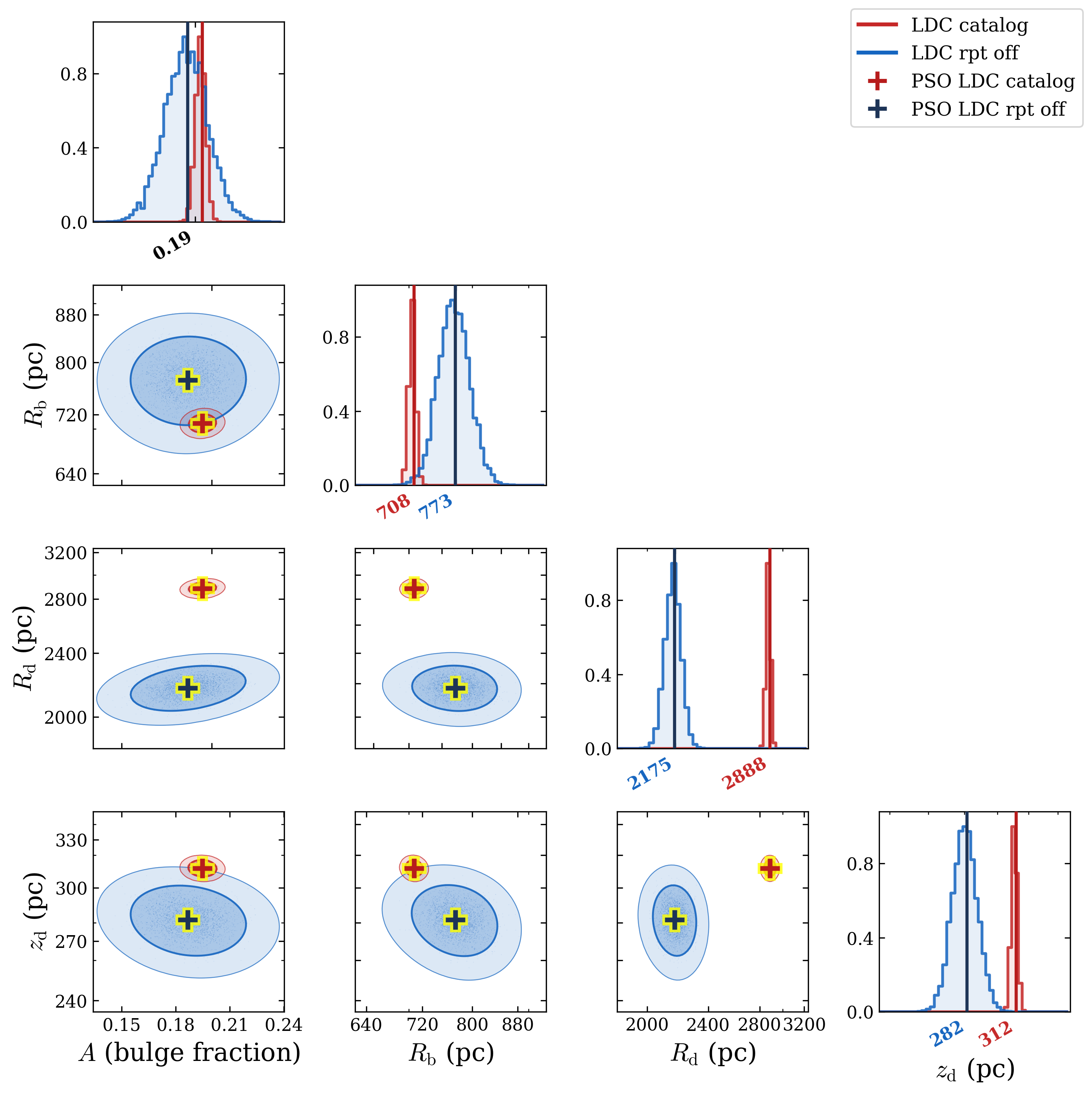}
\caption{Overlay of LDC truth-catalogue posterior (red) and GBSIEVER-reported posterior (blue). Diagonal panels: weighted histograms.  Off-diagonal: resampled scatter with 68\%/95\% kernel-density-estimate contours.  Axis ticks on diagonals mark the weighted medians; when the two medians are close enough that separate labels would overlap, a single tick at their common value (their mean) is drawn instead.  The small cross marks the particle-swarm best fit of the GBSIEVER-reported run; the exact-position truth-catalogue run was not PSO-optimised.}
\label{fig:overlay_truth}
\end{figure}

To isolate the impact of measurement errors, we overlay the GBSIEVER posteriors with results from the LDC truth catalogue, for which source positions are known exactly and the exact-position likelihood of Eq.~\eqref{eq:ipp_exact} applies (run on a $51^3$ grid).  Fig.~\ref{fig:overlay_truth} shows the direct comparison by overlaying the two posteriors, while Table~\ref{tab:gbsiever_catalog_results} provides a quantitative summary.  On the diagonal panels the tick labels report the weighted medians of the two overlaid posteriors; when the two medians are too close to be legibly separated, a single label at their common value (the mean of the two medians) is drawn in place of two overlapping labels.  The diagonal tick values therefore do not in general equal the individual posterior medians quoted in the text, which are always the per-run values of Table~\ref{tab:gbsiever_catalog_results}.

\begin{table}[t]
\caption{Comparison of GBSIEVER single LISA detector-reported source list on LDC data with the LDC truth catalogue.  The truth run uses exact positions and a $51^3$ grid; the GBSIEVER run uses the full $\mathbf{W}$-matrix likelihood on a $101^3$ grid.  Cells give the posterior median with 68\% credible interval.}
\label{tab:gbsiever_catalog_results}
\begin{center}
\renewcommand{\arraystretch}{1.4}
\small\setlength{\tabcolsep}{4pt}
\begin{tabular}{lcccc}\toprule
Param  & LDC (LISA) & LDC catalog \\\midrule
$A$           & $0.187^{+0.011}_{-0.012}$ & $0.195^{+0.003}_{-0.003}$  \\
$R_b$ (pc)    & $773^{+26}_{-25}$  & $708^{+5}_{-5}$  \\
$R_d$ (pc)    & $2175^{+51}_{-52}$ & $2888^{+19}_{-19}$ \\
$z_d$ (pc)    & $282^{+7}_{-7}$  & $312^{+2}_{-2}$  \\\bottomrule
\end{tabular}
\end{center}
\end{table}

The overlay shows the expected behaviour: the GBSIEVER-reported posterior (blue) is markedly broader than the truth-catalogue posterior (red).  With exact positions, the truth run resolves the small-scale structure that pins down the sub-kpc parameters, yielding tight credible intervals ($A=0.195^{+0.003}_{-0.003}$, $R_b=708^{+5}_{-5}$~pc, $z_d=312^{+2}_{-2}$~pc, $R_d=2888^{+19}_{-19}$~pc on the $51^3$ grid).  Propagating the per-source measurement errors through the $\mathbf{W}$ matrix instead smears each source over its position kernel (Fig.~\ref{fig:snr_unc}), washing out this small-scale information and broadening the posteriors by a factor $\sim5$ in $R_b$ and $\sim3.5$ in $z_d$ (to $R_b=773^{+26}_{-25}$~pc, $z_d=282^{+7}_{-7}$~pc) and by a factor $\sim2.7$ in $R_d$ (to $R_d=2175^{+51}_{-52}$~pc on the $101^3$ grid).  The central values also shift---$R_b$ increases while $z_d$ and $R_d$ decrease---reflecting the differences in source lists and selection effects between the GBSIEVER search output and the LDC truth catalogue, rather than being driven by the error convolution alone.

\subsection{Impact of the astrophysical prior}\label{subsec:LDCprior}

The GBSIEVER search incorporates the astrophysical prior of \cite{Zhao:2026dse} into the per-source parameter estimation. This prior places a frequency-dependent constraint on $\dot{f}$, restricting its value to a specified range that scales with $f$. In the absence of this prior, the distance estimator (Eq.~\ref{eq:distance}), which scales as $R\propto\dot{f}/f^{3}\mathcal{A}$, tends to systematically overestimate the frequency derivatives of low-SNR sources near the detection threshold. Such overestimates scatter the inferred distances by $50\%$--$100\%$, pushing sources away from the physically plausible disk/bulge region. By regularising $\dot{f}$ estimates toward the prior-allowed range, the recovery of individual source parameters is measurably improved.

\begin{figure}
\centering
\includegraphics[width=0.48\textwidth]{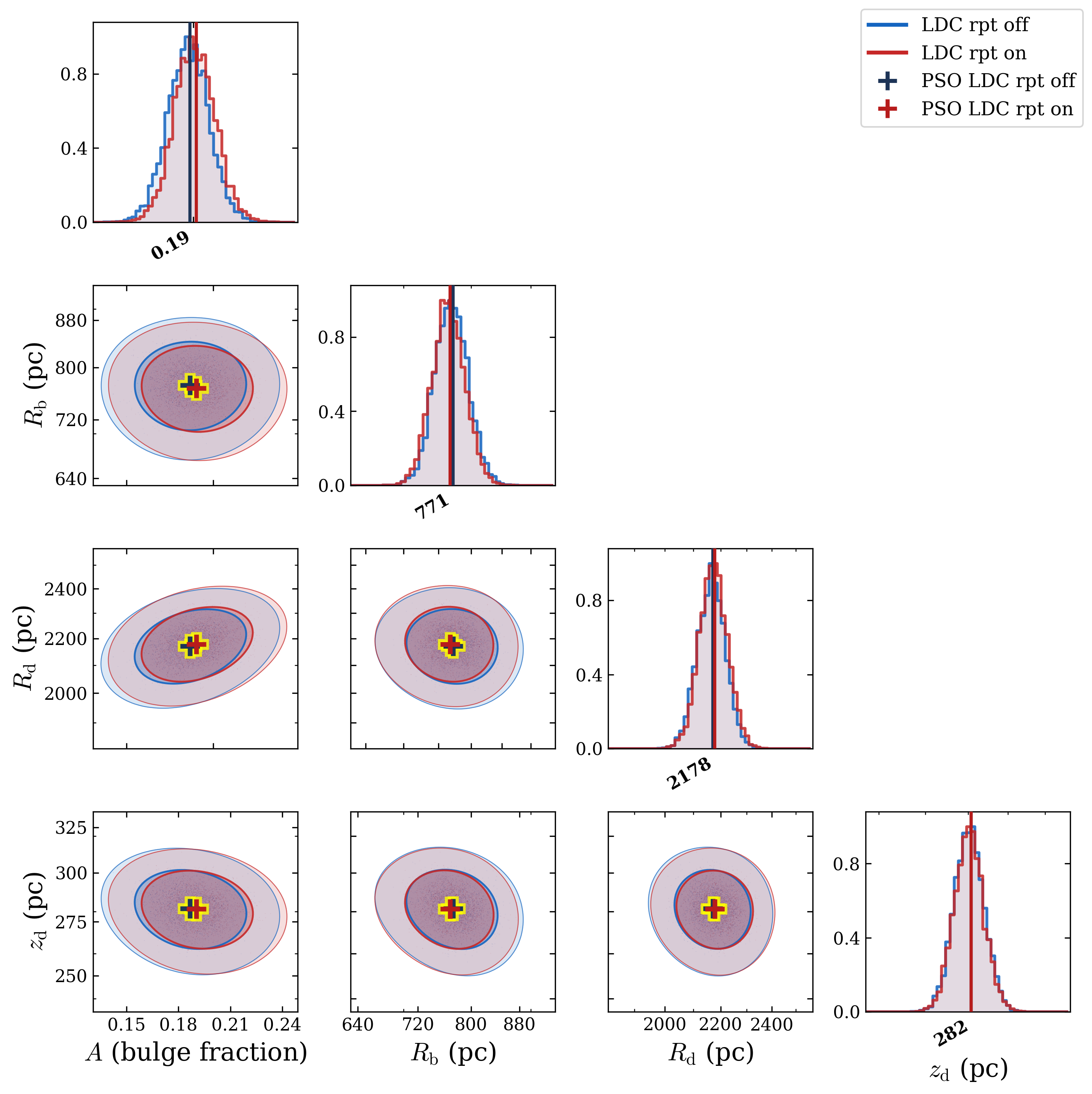}
\caption{Overlay of LDC prior-on (red) vs prior-off (blue) posteriors.  Diagonal ticks follow the same convention as Fig.~\ref{fig:overlay_truth}: when the two medians overlap a single tick at their common value is drawn, otherwise two coloured ticks mark the individual medians.  The small crosses mark the particle-swarm best fits of the two runs.}
\label{fig:overlay_onoff}
\end{figure}

The improvement does not survive our selection, however.  The $\dot{f}>1/T_{\rm obs}^{2}$ cut already removes most of the low-$\dot{f}$ sources that the prior regularises---indeed, more prior-on sources fail this cut ($3021$ survive it versus $3559$ prior off), because the prior moves some overestimated $\dot{f}$ values below the threshold---and the subsequent $R<15$~kpc and NN$<500$~pc cuts leave $2623$ and $2604$ sources, respectively, converging to $N_{\rm det}=2175$ versus $2151$ after all cuts, a difference of $\sim1\%$.  The resulting posteriors are statistically indistinguishable (Fig.~\ref{fig:overlay_onoff} and Table~\ref{tab:gbsiever_results}): the Fisher-based values are $A=0.190^{+0.012}_{-0.011}$ versus $0.187^{+0.011}_{-0.012}$, $R_b=769^{+25}_{-24}$ versus $773^{+26}_{-25}$~pc, $R_d=2181^{+53}_{-50}$ versus $2175^{+51}_{-52}$~pc, and $z_d=282^{+7}_{-7}$ versus $282^{+7}_{-7}$~pc (prior on versus off).  The astrophysical prior therefore improves the per-source parameter estimation at the search stage, but for the sources that pass our quality cuts its impact on the hierarchical inference is negligible.

\subsection{Grid-resolution convergence}\label{subsec:grid}

\begin{table*}[ht]
\caption{Grid-resolution convergence of the GBSIEVER-reported LDC (prior off) posterior as a function of integration-grid resolution.  Cells give the posterior median with 68\% credible interval.  The sub-kpc parameters $R_b$ and $z_d$ converge once $\Delta x\lesssim300$~pc and $\Delta z\lesssim80$~pc, and the $51^3$, $101^3$, and $151^3$ results agree within the credible intervals, confirming that the $101^3$ grid is already converged.}
\label{tab:grid_conv}
\begin{center}
\renewcommand{\arraystretch}{1.4}
\setlength{\tabcolsep}{4pt}
\begin{tabular}{lccccc}\toprule
Param  & $21^3$ & $31^3$ & $51^3$ & $101^3$ & $151^3$\\\midrule
$A$  & $0.149^{+0.013}_{-0.013}$ & $0.173^{+0.012}_{-0.013}$ & $0.184^{+0.012}_{-0.012}$ & $0.187^{+0.012}_{-0.011}$ & $0.187^{+0.012}_{-0.012}$\\
$R_b$ (pc)  & $1062.9^{+31.8}_{-34.7}$    & $874.9^{+27.3}_{-29.7}$     & $793.1^{+26.7}_{-26.9}$        & $773.1^{+24.6}_{-25.6}$ & $772.0^{+24.4}_{-25.9}$\\
$R_d$ (pc)  & $2173.9^{+51.0}_{-53.9}$    & $2116.4^{+49.5}_{-54.5}$     & $2155.4^{+51.8}_{-50.3}$     & $2175.0^{+51.9}_{-50.6}$ & $2187.0^{+50.6}_{-53.8}$\\
$z_d$ (pc)    & $338.1^{+8.0}_{-8.7}$      & $311.1^{+8.0}_{-7.9}$        & $284.6^{+7.2}_{-7.5}$        & $281.7^{+7.1}_{-7.4}$ & $279.4^{+6.9}_{-7.0}$\\
\bottomrule
\end{tabular}
\end{center}
\end{table*}

Table~\ref{tab:grid_conv} summarises the GBSIEVER-reported LDC (prior off) posterior as a function of integration-grid resolution.  The sub-kpc parameters are the most resolution-sensitive: $R_b$ falls from $1063$~pc at $21^3$ to $773^{+26}_{-25}$~pc at $101^3$, and $z_d$ from $338$~pc to $282^{+7}_{-7}$~pc, converging once $\Delta x\lesssim300$~pc and $\Delta z\lesssim80$~pc.  The bulge fraction $A$ rises monotonically but mildly from $0.149$ to $0.187$, while the kpc-scale $R_d$ is essentially resolution-independent, varying only within $2116$--$2187$~pc across the five grids, non-monotonically and within the credible intervals.  The $51^3$, $101^3$, and $151^3$ posteriors agree within their mutual 68\% intervals for every parameter, and between $101^3$ and $151^3$ every posterior median shifts by $\lesssim1\%$ ($R_d=2175.0$ versus $2187.0$~pc, $z_d=281.7$ versus $279.4$~pc, $R_b=773.1$ versus $772.0$~pc), confirming that the $101^3$ grid of the main results is already converged.

\section{Discussion}

The Fisher-based results place the recovered disk parameters in good agreement with the canonical Galactic values.  The scale height $z_d=282^{+7}_{-7}$~pc is consistent with the $\sim300$~pc measured for the stellar thin disk (Table~\ref{tab:gbsiever_results}) and with the $z_d\approx312$~pc recovered from the exact-position truth-catalogue fit, indicating that the waveform-Fisher error convolution no longer biases the vertical structure.  The scale length $R_d=2175^{+51}_{-52}$~pc is $\sim13\%$ below the canonical $2500$~pc.  The bulge fraction $A=0.187^{+0.011}_{-0.012}$ is now measured at $\sim6\%$ statistical precision, much closer to the canonical value of $0.25$ than in our earlier exact-position analysis~\cite{universe11080248}.

The catalogue-to-catalogue scatter is a useful diagnostic of systematic error.  The LDC and TDC searches are built from different data streams and source lists, yet they recover statistically consistent scale heights ($z_d=282$ versus $276$~pc) and scale lengths ($R_d=2175$ versus $2234$~pc); the larger differences in the bulge parameters ($A=0.187$ versus $0.222$ and $R_b=773$ versus $700$~pc) reflect the weaker information content of the bulge, which is sampled by far fewer sources than the disk.  This scatter, rather than the formal $\sim3$--$6\%$ statistical uncertainties, is the more honest estimate of the current systematic floor.

\section{Conclusions and Outlook}
\label{sec:conclusions}

We have presented \textsc{Galena}, a hierarchical Bayesian pipeline that infers Milky Way structural parameters from resolved double-white-dwarf gravitational-wave sources through an IPP likelihood.  Per-source position covariances are propagated analytically via a $3\times5$ Jacobian and folded into a pre-computed sparse $\mathbf{W}$-matrix, yielding an exact-position path for truth catalogues and a $\mathbf{W}$-matrix path for search outputs that incorporates position uncertainties, the main limitation of our earlier exact-position maximum-likelihood approach~\cite{universe11080248}.

Applied to GBSIEVER-reported LDC sources ($N=2151$ after quality cuts), \textsc{Galena} constrains the structural parameters to $R_b=773^{+26}_{-25}$~pc, $R_d=2175^{+51}_{-52}$~pc, $z_d=282^{+7}_{-7}$~pc and $A=0.187^{+0.011}_{-0.012}$ on a $101^3$ grid.  The disk scale parameters agree with canonical Milky Way values, with $z_d$ consistent with the canonical thin-disk scale height, whereas the bulge parameters $A$ and $R_b$ remain weakly constrained.  The search-stage astrophysical prior improves the per-source $\dot{f}$ estimates but leaves the final inference essentially unchanged, and a joint LISA+Taiji-like run shifts the posterior while narrowing $R_b$ only modestly.  The particle-swarm best fits, marked by the crosses in Figs.~\ref{fig:corner_ldc} and \ref{fig:corner_tdc}, fall on the posterior peaks, confirming that these constraints do not depend on the choice of sampler.

The current analysis has three main limitations.  First, the two-component density model omits the thick disk and stellar halo \cite{tkachenko2025determining}, whose DWD populations contribute non-negligibly at large scale heights; adding these components would demand a correspondingly larger catalogue to avoid degeneracies.  Second, the per-source position covariances are built from the Fisher-matrix (Gaussian) approximation to each source's waveform likelihood rather than the full non-Gaussian posterior, so the parameter correlations shaping $\mathbf{C}_{\mathbf{x}}^{(i)}$ are captured only to second order.  Third, the hard distance cut at $15$~kpc is a placeholder for the true LISA selection function, which depends on chirp mass, frequency, and sky position in addition to distance \cite{Korol2022}.  Of these, the selection function is the most consequential: the hard-sphere cut suffices to prevent the unphysical $R_d\to\infty$ of uncorrected fits, but a realistic SNR-dependent $\bar{\eta}(\mathcal{M}_{c},f,R)$ from injection-recovery simulations would break the geometric degeneracy between $R_d$ and the selection boundary.

The joint LISA+Taiji-like run narrows the $R_b$ credible interval ($^{+26}_{-25}\to^{+22}_{-21}$~pc) while leaving $R_d$ and $z_d$ essentially unchanged, yet the gain remains limited: the strict cuts retain essentially the same sources as LISA-only, and the Fisher-matrix error propagation captures only part of the two-detector per-source improvement.  Relaxing the cuts is the natural next step toward realising the full network gain anticipated for space-borne detectors \cite{zhang2022resolving}.  The analytical Jacobian has been verified against finite-difference derivatives to $10^{-6}$ relative error.

The analytical Jacobian and sparse $\mathbf{W}$-matrix deliver $\sim5$ likelihood evaluations per second at the $101^{3}$ resolution of the main results on a single core (up to $\sim700$ at $21^{3}$), and a full nested-sampling run completes in a few hours at $101^{3}$; the pipeline scales linearly with $N_{\rm det}$ for likelihood evaluation and quadratically for the one-time $\mathbf{W}$ construction, so catalogues of $10^{4}$--$10^{5}$ sources are within reach with modest parallelisation.  Grid-convergence tests show that sub-kpc parameters require $\Delta x\lesssim300$~pc and $\Delta z\lesssim80$~pc for unbiased recovery.  

Future work will add a realistic SNR-dependent selection function, a thick-disk and halo component, a self-consistent joint fit of the DWD population-synthesis and Galactic structure parameters, a joint LISA--Taiji-like re-analysis with relaxed quality cuts, and---once a public TianQin mock-data catalogue or data challenge becomes available---an application of the same pipeline to TianQin data and to LISA--Taiji--TianQin network configurations.

\section*{Acknowledgments}

This study was supported by the National Key Research and Development Program of China (Grant No.~2023YFC2206701 and No.~2021YFC2203003), the National Natural Science Foundation of China (Grants No.~12475056, No.~12247101), the Fundamental Research Funds for the Central Universities (Grant No.~lzujbky-2025-jdzx07), the Natural Science Foundation of Gansu Province (No.~22JR5RA389, No.~25JRRA799), the 111 Project (Grant No.~B20063) and Gansu Province's Top Leading Talent Support Plan.

\section*{Data Availability}

The LISA Data Challenge (\href{https://lisa-ldc.in2p3.fr/}{LDC1-4 GB v2}) and Taiji Data Challenge (\href{http://taiji-tdc.ictp-ap.org/dataset_page/}{time domain data}) used in this work are publicly available; the GBSIEVER search outputs and the \textsc{Galena} pipeline code are available from the corresponding author upon reasonable request.

\bibliographystyle{apsrev4-2}
\bibliography{refs}

@article{AmaroSeoane2017,
  title={Laser interferometer space antenna},
  author={Amaro-Seoane, Pau and Audley, Heather and Babak, Stanislav and Baker, John and Barausse, Enrico and Bender, Peter and Berti, Emanuele and Binetruy, Pierre and Born, Michael and Bortoluzzi, Daniele and others},
  journal={arXiv preprint arXiv:1702.00786},
  year={2017},
  eprint  = {1702.00786},
}

@article{Hu2017,
  title={The Taiji Program in Space for gravitational wave physics and the nature of gravity},
  author={Hu, Wen-Rui and Wu, Yue-Liang},
  journal={National Science Review},
  volume={4},
  number={5},
  pages={685--686},
  year={2017},
  publisher={Oxford University Press}
}

@Article{universe11080248,
AUTHOR = {Zhao, Shao-Dong and Zhang, Xue-Hao and Mohanty, Soumya D. and Fullana i Alfonso, Màrius Josep and Liu, Yu-Xiao and Xie, Qun-Ying},
TITLE = {Estimating Galactic Structure Using Galactic Binaries Resolved by Space-Based Gravitational Wave Observatories},
JOURNAL = {Universe},
VOLUME = {11},
YEAR = {2025},
NUMBER = {8},
ARTICLE-NUMBER = {248},
URL = {https://www.mdpi.com/2218-1997/11/8/248},
ISSN = {2218-1997},
DOI = {10.3390/universe11080248}
}

@article{Zhao:2026dse,
  title={Improving the resolution of double white dwarf systems with spaceborne gravitational wave observatories using a robust astrophysical prior},
  author={Zhao, Shao-Dong and Zhang, Xue-Hao and Mohanty, Soumya D and Liu, Yu-Xiao},
  journal={arXiv preprint arXiv:2606.01236},
  year={2026}
}

@article{Luo2016,
  title={TianQin: a space-borne gravitational wave detector},
  author={Luo, Jun and Chen, Li-Sheng and Duan, Hui-Zong and Gong, Yun-Gui and Hu, Shoucun and Ji, Jianghui and Liu, Qi and Mei, Jianwei and Milyukov, Vadim and Sazhin, Mikhail and others},
  journal={Classical and Quantum Gravity},
  volume={33},
  number={3},
  pages={035010},
  year={2016},
  publisher={IOP Publishing}
}

@article{Nelemans2001,
  title={The gravitational wave signal from the Galactic disk population of binaries containing two compact objects},
  author={Nelemans, Gijs and Yungelson, LR and Portegies Zwart, Simon F},
  journal={Astronomy \& Astrophysics},
  volume={375},
  number={3},
  pages={890--898},
  year={2001},
  publisher={EDP Sciences}
}

@article{Ruiter2010,
  title={The LISA gravitational wave foreground: a study of double white dwarfs},
  author={Ruiter, Ashley J and Belczynski, Krzysztof and Benacquista, Matthew and Larson, Shane L and Williams, Gabriel},
  journal={The Astrophysical Journal},
  volume={717},
  number={2},
  pages={1006--1021},
  year={2010},
  publisher={The American Astronomical Society}
}

@article{Korol2022,
  title={Observationally driven Galactic double white dwarf population for LISA},
  author={Korol, Valeriya and Hallakoun, Na’ama and Toonen, Silvia and Karnesis, Nikolaos},
  journal={Monthly Notices of the Royal Astronomical Society},
  volume={511},
  number={4},
  pages={5936--5947},
  year={2022},
  publisher={Oxford University Press}
}

@article{Wilhelm:2020qjc,
    author = "Wilhelm, Martijn J. C. and Korol, Valeriya and Rossi, Elena M. and D'Onghia, Elena",
    title = "{The Milky Way's bar structural properties from gravitational waves}",
    eprint = "2003.11074",
    archivePrefix = "arXiv",
    primaryClass = "astro-ph.GA",
    doi = "10.1093/mnras/staa3457",
    journal = "Mon. Not. Roy. Astron. Soc.",
    volume = "500",
    number = "4",
    pages = "4958--4971",
    year = "2020"
}

@misc{LDC2021guide,
  author  = {{LISA Data Challenge Working Group}},
  title   = {LDC1-4 GB v2 Catalogue Guide},
  year    = {2021},
  url     = {https://lisa-ldc.lal.in2p3.fr},
}

@article{Littenberg2011,
  title={Detection pipeline for Galactic binaries in LISA data},
  author={Littenberg, Tyson B},
  journal={Physical Review D—Particles, Fields, Gravitation, and Cosmology},
  volume={84},
  number={6},
  pages={063009},
  year={2011},
  publisher={APS}
}

@article{Farr2019,
  title={Accuracy requirements for empirically-measured selection functions},
  author={Farr, Will M},
  journal={arXiv preprint arXiv:1904.10879},
  year={2019}
}

@article{Mandel2019,
  title={Extracting distribution parameters from multiple uncertain observations with selection biases},
  author={Mandel, Ilya and Farr, Will M and Gair, Jonathan R},
  journal={Monthly Notices of the Royal Astronomical Society},
  volume={486},
  number={1},
  pages={1086--1093},
  year={2019},
  publisher={Oxford University Press}
}

@article{BlandHawthorn2016,
  title={The galaxy in context: structural, kinematic, and integrated properties},
  author={Bland-Hawthorn, Joss and Gerhard, Ortwin},
  journal={Annual Review of Astronomy and Astrophysics},
  volume={54},
  pages={529--596},
  year={2016},
  publisher={Annual Reviews}
}

@article{Bovy2012,
  title={The Milky Way's circular-velocity curve between 4 and 14 kpc from APOGEE data},
  author={Bovy, Jo and Allende Prieto, Carlos and Beers, Timothy C and Bizyaev, Dmitry and Da Costa, Luiz N and Cunha, Katia and Ebelke, Garrett L and Eisenstein, Daniel J and Frinchaboy, Peter M and Garc{\'\i}a P{\'e}rez, Ana Elia and others},
  journal={The Astrophysical Journal},
  volume={759},
  number={2},
  pages={131},
  year={2012},
  publisher={The American Astronomical Society}
}

@article{Juric2008,
  title={The Milky Way tomography with SDSS. I. Stellar number density distribution},
  author={Juri{\'c}, Mario and Ivezi{\'c}, {\v{Z}}eljko and Brooks, Alyson and Lupton, Robert H and Schlegel, David and Finkbeiner, Douglas and Padmanabhan, Nikhil and Bond, Nicholas and Sesar, Branimir and Rockosi, Constance M and others},
  journal={The Astrophysical Journal},
  volume={673},
  number={2},
  pages={864--914},
  year={2008}
}

@article{HAYNSWORTH196873,
title = {Determination of the inertia of a partitioned Hermitian matrix},
journal = {Linear Algebra and its Applications},
volume = {1},
number = {1},
pages = {73-81},
year = {1968},
issn = {0024-3795},
doi = {https://doi.org/10.1016/0024-3795(68)90050-5},
url = {https://www.sciencedirect.com/science/article/pii/0024379568900505},
author = {Emilie V. Haynsworth}
}

@article{Speagle2020,
  title={dynesty: a dynamic nested sampling package for estimating Bayesian posteriors and evidences},
  author={Speagle, Joshua S},
  journal={Monthly Notices of the Royal Astronomical Society},
  volume={493},
  number={3},
  pages={3132--3158},
  year={2020},
  publisher={Oxford University Press}
}

@article{breivik2020constraining,
  title={Constraining galactic structure with the LISA white dwarf foreground},
  author={Breivik, Katelyn and Mingarelli, Chiara MF and Larson, Shane L},
  journal={The Astrophysical Journal},
  volume={901},
  number={1},
  pages={4},
  year={2020},
  publisher={The American Astronomical Society}
}

@article{korol2019multimessenger,
  title={A multimessenger study of the Milky Way’s stellar disc and bulge with LISA, Gaia, and LSST},
  author={Korol, Valeriya and Rossi, Elena M and Barausse, Enrico},
  journal={Monthly Notices of the Royal Astronomical Society},
  volume={483},
  number={4},
  pages={5518--5533},
  year={2019},
  publisher={Oxford University Press}
}

@article{georgousi2023gravitational,
  title={Gravitational waves from double white dwarfs as probes of the milky way},
  author={Georgousi, Maria and Karnesis, Nikolaos and Korol, Valeriya and Pieroni, Mauro and Stergioulas, Nikolaos},
  journal={Monthly Notices of the Royal Astronomical Society},
  volume={519},
  number={2},
  pages={2552--2566},
  year={2023},
  publisher={Oxford University Press}
}

@article{Zhang2025vya,
    author = "Zhang, Xue-Hao and Mohanty, Soumya D. and Valluri, S. R. and Zhao, Shao-Dong and Xie, Qun-Ying and Liu, Yu-Xiao",
    title = "{Efficient Parallel Processing of Second-Generation TDI Data for Galactic Binaries in Space-Based Gravitational Wave Missions}",
    doi = "10.3390/universe11090313",
    journal = "Universe",
    volume = "11",
    number = "9",
    pages = "313",
    year = "2025"
}

@article{zhang2024constraining,
  title={Constraining the Galactic Structure Using Time Domain Gravitational Wave Signal from Double White Dwarfs Detected by Space Gravitational Wave Detectors},
  author={Zhang, Siqi and Deng, Furen and Lu, Youjun and Yu, Shenghua},
  journal={The Astrophysical Journal},
  volume={978},
  number={1},
  pages={61},
  year={2025},
  publisher={The American Astronomical Society}
}

@article{korol2020populations,
  title={Populations of double white dwarfs in Milky Way satellites and their detectability with LISA},
  author={Korol, V and Toonen, S and Klein, A and Belokurov, V and Vincenzo, F and Buscicchio, R and Gerosa, D and Moore, CJ and Roebber, E and Rossi, EM and others},
  journal={Astronomy \& Astrophysics},
  volume={638},
  pages={A153},
  year={2020},
  publisher={EDP Sciences}
}

@article{ruan2020taiji,
  title={Taiji program: Gravitational-wave sources},
  author={Ruan, Wen-Hong and Guo, Zong-Kuan and Cai, Rong-Gen and Zhang, Yuan-Zhong},
  journal={International Journal of Modern Physics A},
  volume={35},
  number={17},
  pages={2050075},
  year={2020},
  publisher={World Scientific}
}

@article{karnesis2021stochastic,
  title={Characterization of the stochastic signal originating from compact binary populations as measured by LISA},
  author={Karnesis, Nikolaos and Babak, Stanislav and Pieroni, Mauro and Cornish, Neil and Littenberg, Tyson},
  journal={Physical Review D},
  volume={104},
  number={4},
  pages={043019},
  year={2021},
  publisher={APS}
}

@article{zhang2021pso,
  title={Resolving Galactic binaries in LISA data using particle swarm optimization and cross-validation},
  author={Zhang, Xue-Hao and Mohanty, Soumya D and Zou, Xiao-Bo and Liu, Yu-Xiao},
  journal={Physical Review D},
  volume={104},
  number={2},
  pages={024023},
  year={2021},
  publisher={APS}
}

@article{zhang2022resolving,
  title={Resolving Galactic binaries using a network of space-borne gravitational wave detectors},
  author={Zhang, Xue-Hao and Zhao, Shao-Dong and Mohanty, Soumya D and Liu, Yu-Xiao},
  journal={Physical Review D},
  volume={106},
  number={10},
  pages={102004},
  year={2022},
  publisher={APS}
}

@inproceedings{eberhart1995particle,
  title="{Particle swarm optimization}",
  author={Eberhart, Russell and Kennedy, James},
  booktitle={Proceedings of the IEEE international conference on neural networks},
  volume={4},
  pages={1942--1948},
  year={1995},
  organization={Citeseer}
}

@article{babak2010mock,
  title={The mock LISA data challenges: from challenge 3 to challenge 4},
  author={Babak, Stanislav and Baker, John G and Benacquista, Matthew J and Cornish, Neil J and Larson, Shane L and Mandel, Ilya and McWilliams, Sean T and Petiteau, Antoine and Porter, Edward K and Robinson, Emma L and others},
  journal={Classical and Quantum Gravity},
  volume={27},
  number={8},
  pages={084009},
  year={2010},
  publisher={IOP Publishing}
}

@article{baghi2022lisa,
  author  = {Baghi, Quentin},
  title   = {The LISA Data Challenges},
  journal = {arXiv preprint arXiv:2204.12142},
  year    = {2022},
}

@misc{babak2020lisa,
    author = {Babak, Stanislav and Petiteau, Antoine},
    title = "{LISA Data Challenge Manual}",
    year = 2020,
    howpublished = {\url{https://lisa-ldc.lal.in2p3.fr/static/data/pdf/LDC-manual-002.pdf}},
}

@book{kay1993fundamentals,
  title={Fundamentals of statistical signal processing: estimation theory},
  author={Kay, Steven M},
  year={1993},
  publisher={Prentice-Hall, Inc.}
}

@article{balasubramanian1996gravitational,
  title={Gravitational waves from coalescing binaries: Detection strategies and Monte Carlo estimation of parameters},
  author={Balasubramanian, Ramachandran and Sathyaprakash, Bangalore Suryanarayana and Dhurandhar, SV},
  journal={Physical Review D},
  volume={53},
  number={6},
  pages={3033},
  year={1996},
  publisher={APS}
}

@article{Jaranowski1998,
  title={Data analysis of gravitational-wave signals from spinning neutron stars: The signal and its detection},
  author={Jaranowski, Piotr and Krolak, Andrzej and Schutz, Bernard F},
  journal={Physical Review D},
  volume={58},
  number={6},
  pages={063001},
  year={1998},
  publisher={APS}
}

@article{tinto2014time,
  title="{Time-delay interferometry}",
  author={Tinto, Massimo and Dhurandhar, Sanjeev V},
  journal={Living Reviews in Relativity},
  volume={17},
  number={1},
  pages={1--54},
  year={2014},
  publisher={Springer}
}

@article{adams2012astrophysical,
  title={Astrophysical model selection in gravitational wave astronomy},
  author={Adams, Matthew R and Cornish, Neil J and Littenberg, Tyson B},
  journal={Physical Review D},
  volume={86},
  number={12},
  pages={124032},
  year={2012},
  publisher={APS}
}

@article{reid2004proper,
  title={The proper motion of Sagittarius A*. II. The mass of Sagittarius A},
  author={Reid, Mark J and Brunthaler, A},
  journal={The Astrophysical Journal},
  volume={616},
  number={2},
  pages={872},
  year={2004},
  publisher={IOP Publishing}
}

@article{reid2014trigonometric,
  title={Trigonometric parallaxes of high mass star forming regions: the structure and kinematics of the Milky Way},
  author={Reid, Mark J and Menten, Karl M and Brunthaler, Andreas and Zheng, Xing-Wu and Dame, Thomas M and Xu, Ye and Wu, Ya and Zhang, B and Sanna, Andrea and Sato, Makoto and others},
  journal={The Astrophysical Journal},
  volume={783},
  number={2},
  pages={130},
  year={2014},
  publisher={The American Astronomical Society}
}

@article{abuter2019geometric,
  title={A geometric distance measurement to the Galactic center black hole with 0.3\% uncertainty},
  author={Abuter, Roberto and Amorim, A and Baub{\"o}ck, M and Berger, JP and Bonnet, H and Brandner, W and Cl{\'e}net, Y and Du Foresto, V Coud{\'e} and De Zeeuw, PT and Dexter, J and others},
  journal={Astronomy \& Astrophysics},
  volume={625},
  pages={L10},
  year={2019},
  publisher={EDP Sciences}
}

@article{bennett2019vertical,
  title={Vertical waves in the solar neighbourhood in Gaia DR2},
  author={Bennett, Morgan and Bovy, Jo},
  journal={Monthly Notices of the Royal Astronomical Society},
  volume={482},
  number={1},
  pages={1417--1425},
  year={2019},
  publisher={Oxford University Press}
}

@article{tkachenko2025determining,
  title={Determining the Scale Length and Height of the Milky Way’s Thick Disc Using RR Lyrae},
  author={Tkachenko, Roman and Vieira, Katherine and Lutsenko, Artem and Korchagin, Vladimir and Carraro, Giovanni},
  journal={Universe},
  volume={11},
  number={4},
  pages={132},
  year={2025},
  publisher={MDPI}
}

\end{document}